\documentclass[12pt]{article}
\usepackage{times}
\usepackage{ametsoc}
\usepackage{bm}

\newcommand{\be}{\begin{equation}}
\newcommand{\ee}{\end{equation}}
\newcommand{\bes}{\begin{equation*}}
\newcommand{\ees}{\end{equation*}}
\newcommand{\bdm}{\begin{displaymath}}
\newcommand{\edm}{\end{displaymath}}
\newcommand{\ba}{\begin{align}}
\newcommand{\ea}{\end{align}}
\newcommand{\bas}{\begin{align*}}
\newcommand{\eas}{\end{align*}}

\newcommand{\eqref}[1]{(\ref{#1})}

\newcommand{\dg}{\ensuremath{^{\circ}}}

\def\beqn{\begin{eqnarray}}
\def\eeqn{\end{eqnarray}}
\newcommand{\benu}{\begin{enumerate}}
\newcommand{\eenu}{\end{enumerate}}
\newcommand{\bdes}{\begin{description}}
\newcommand{\edes}{\end{description}}
\newcommand{\bfig}{\begin{figure}}
\newcommand{\efig}{\end{figure}}
\newcommand{\bcen}{\begin{center}}
\newcommand{\ecen}{\end{center}}
\newcommand{\beq}{\begin{equation}}
\newcommand{\eeq}{\end{equation}}
 \usepackage{graphicx}

\begin{document}
\begin{titlepage}

\begin{center}
{\Large\bf  Surface fluxes and tropical intraseasonal variability:  a reassessment}
\vspace{1.5cm}
                                                     
ADAM H. SOBEL \footnote{ \baselineskip 3mm \it Corresponding Author
Address: \rm Adam Sobel, Columbia University, Dept. of Applied Physics
and Applied Mathematics, 500 W. 120th St., Rm. 217, New York, NY 10027 USA. 
E-mail: ahs129@columbia.edu}
\vspace{.2cm}
                                                                                                   
{\it Department of Applied Physics and Applied Mathematics and Department of Earth and Environmental Sciences \\ Columbia University, New York, NY},
\rm

\vspace{1.0cm}

ERIC D. MALONEY
\vspace{.2cm}

{\it Department of Atmospheric Sciences\\Colorado State University, Fort Collins, CO},
\rm
\vspace{1.0cm}

GILLES BELLON
\vspace{.2cm}

{\it Department of Applied Physics and Applied Mathematics \\ Columbia University, New York, NY},
\rm
\vspace{1.0cm}

DARGAN M. FRIERSON
\vspace{.2cm}

{\it Department of Atmospheric Sciences\\University of Washington, Seattle, WA.}
\rm
\vspace{1.0cm}

\normalsize {\em May 10, 2008}
\end{center}
\end{titlepage}

\begin{abstract}
The authors argue for the hypothesis that interactive feedbacks
involving surface enthalpy fluxes are important to the dynamics of
tropical intraseasonal variability.  These include
cloud-radiative feedbacks as well as surface turbulent flux
feedbacks;  the former effectively
act to transport enthalpy from the ocean to the atmosphere, as do the
latter.  Evidence in favor of this hypothesis includes the observed 
spatial distribution of intraseasonal variance in
precipitation and outgoing longwave radiation, the observed relationship
between intraseasonal latent heat flux and precipitation anomalies in
regions where intraseasonal variability is strong, and sensitivity 
experiments performed with a small number of general circulation
and idealized models.

The authors argue that it would be useful to assess
the importance of surface fluxes to intraseasonal variability in a larger
number of comprehensive numerical
models.  Such an assessment could provide insight into the
relevance of interactive surface fluxes to real intraseasonal variability,
perhaps making it possible to rule out either theoretical explanations 
in which surface fluxes are crucial,
or those in which they are not.

\end{abstract}

\clearpage

\section{Introduction}

Theoretical understanding of the mechanisms responsible for tropical
intraseasonal variability is limited.  There are many interesting
and plausible ideas in the literature, but there is no agreement on which of
them, if any, is correct.  The Madden-Julian oscillation (MJO) in particular
is arguably the most significant mode of atmospheric variability at any sub-decadal
time scale whose essential features --- its existence, energetics, spatial 
and temporal scales --- remain so unsatisfactorily explained.

In this study, we use the phrase "MJO" to refer to the eastward-propagating 30-60
day mode which is dominant in southern hemisphere summer.  This mode remains present
in northern hemisphere summer, but northern summer also features a northward-propagating
mode, manifest in northward-propagating rain bands over the Indian subcontinent
and adjacent oceans.  We refer to these two modes collectively as "tropical intraseasonal
variability" and treat them to some degree as one phenomenon.  We recognize that
the eastward- and northward-propagating modes have some significant differences, but
argue here that there may be fundamental similarities in their energetics.  

General circulation models (GCMs) simulate tropical intraseasonal variability
with varying degrees of fidelity.  A couple of recent intercomparison studies 
\citep{linet06,zhaet06} show that even the best models still have significant flaws in their MJO simulations.  At the same time, some members of the current generation of models show considerable improvement over previous generations.  While improving GCM simulations is sometimes cited as a motivation for theoretical research into the MJO, no
clear relationship exists between the fidelity of GCM simulations and the state of theoretical understanding, perhaps because of the very different levels of complexity of GCMs as compared to the idealized models used by theorists.  It is not clear that recent improvements in MJO simulation owe anything to theoretical understanding of the mechanism of the MJO.  GCM improvements in MJO simulation often seem to be accidental by-products of broader model development efforts, or results of trial-and-error tuning, or perhaps tuning guided by broader principles not specific to the MJO.  For example, any model change which tends to inhibit deep convection tends to increase variability at all timescales, including the intraseasonal timescale.  This constitutes improvement for models in which
intraseasonal variability is too weak, as is the case with many.

Any steps we can take to narrow
the range of mechanisms which are considered possible explanations
for tropical intraseasonal variability would be valuable, particularly
if some mechanisms can be eliminated convincingly enough to focus
the attention of the community on evaluating the remainder.  
We argue that models of the MJO
in which variations in net surface enthalpy fluxes are crucial are
more likely to be correct than those in which such variations are
unimportant, while recognizing that the pioneering studies which 
first proposed this idea (Emanuel 1987;  Neelin et al. 1987) have proven
incorrect in their details.  Those details are all inessential to the hypothesis
that surface enthalpy fluxes are important to tropical intraseasonal variability.  

Three primary pieces of evidence support our argument:
\begin{enumerate}
\item Intraseasonal variance in precipitation and outgoing longwave
radiation is observed to be larger over ocean than land.  This is true
in both hemispheres, even in regions
where the climatological mean precipitation is larger
over land than ocean.  Since the net surface enthalpy flux must vanish over land,
the land-sea contrast in intraseasonal variance is consistent with 
a role for that flux in generating the variance.
\item Over the oceanic regions of largest intraseasonal variability, 
intraseasonal variations in net surface enthalpy flux and precipitation are 
correlated.  
\item In several general circulation models (as well as some idealized models)
surface enthalpy fluxes are demonstrably important to the simulated
intraseasonal variability.  Experiments with these models suggest 
that the role of surface fluxes is larger in those
models whose MJO simulation is better.
\end{enumerate}
While most of this evidence is not new, the GCM results in particular have
started to become more convincing, partly because the state of the art in 
GCM simulations of intraseasonal variability has improved (e.g., Zhang et al. 2006).  
This, in conjunction with continuing observational and theoretical 
work, has led us to the position
that the case for the importance of surface enthalpy fluxes to observed 
intraseasonal variability has become stronger than it was a decade ago,
and deserves to be systematically re-examined.

In the following section, we provide a highly selective review of some
results and ideas, mostly from the theoretical and modeling literature,
which are relevant to the hypothesis that surface
enthalpy fluxes are important to the dynamics of tropical intraseasonal
variability.  In section 3,
we review some observational results which are consistent with this
hypothesis.  This is followed in section 4 by a discussion of GCM results,
including a presentation
of new results from the NOAA GFDL AM2 model which show that the 
hypothesis appears to have merit in that model, consistent with the
results of Maloney and Sobel (2004) who used a version of the NCAR
model.  In section
5, we discuss the implications 
of these results and propose avenues for further research.

\section{Theory and Modeling:  An Unbalanced Review}
\label{sec:review}
\subsection{Southern summer intraseasonal variability (MJO):
Models of Emanuel (1987) and Neelin et al. (1987)}

We do not attempt to provide a comprehensive or balanced review of MJO theory.
This is beyond our intended
scope, and recent reviews have been done by~\cite{wang_05}
and~\cite{zhang_05}.  Instead we focus on theories in which variations in 
surface enthalpy fluxes figure prominently.  

Approximately simultaneously, \cite{ema87} and \cite{neeet87}
proposed that air-sea interaction could destabilize a moist Kelvin wave,
leading to intraseasonal variability in the tropics.  The arguments
involved linear analysis of idealized moist models in which the temperature structure is
assumed to be represented by a first baroclinic mode, and convection is controlled
by quasi-equilibrium principles.  Essentially, convection acts in these systems
to eliminate some local measure of stability of the column to deep convection.  The
atmosphere is adjusted by convection towards a reference value of
the stability
measure [e.g., convective available potential energy (CAPE)], 
which for present purposes may be assumed zero.  The
dynamics of such models is discussed in more detail in a number of
reviews \nocite{emaet94,smith_97,stevens_et_al_97,emanuel_07,ara04}
 (e.g., Emanuel et al. 1994; Emanuel 1997;  Neelin
1997, Stevens et al. 1996;  Smith 1997;  Arakawa 2004;  Emanuel 2007).  
\cite{neeet87} also performed
numerical simulations with a general circulation model (GCM).

We refer to models of the type discussed by Emanuel et al. (1994) as Òquasi-equilibriumÓ
models. In its most general sense, the term
QE refers to a broader category, including all models in which the
convection is assumed close to statistical equilibrium with its forcings. Traditional QE models incorporate
additional assumptions, in particular that of a pure first baroclinic mode vertical structure,
which can be relaxed without relaxing the assumption of
QE per se. 
In models assuming a first baroclinic mode structure as well as QE, the interaction of
deep convection with large-scale dynamics alone does not generate unstable large-scale
modes.  In the simplest such models 
convection is assumed to respond instantaneously to large-scale forcing (which 
can come from large-scale dynamics, radiation, or surface fluxes)
so as to remove all instability completely; 
Emanuel et al. (1994) called this "strict quasi-equilibrium".
In this case, the interaction of convection and 
large-scale dynamics reduces the effective stratification and thus the phase speed of 
convectively coupled gravity and Kelvin waves, but does not stabilize or destabilize them. If instead
the convection is assumed to relax the CAPE (or other measure 
of instability) towards its reference value with a finite
timescale, then the interaction damps disturbances, a phenomenon known as "moist
convective damping".  Disturbances in these models cannot become linearly unstable
through the interaction of convection with large-scale dynamics alone, but only
through feedbacks involving processes which can act as sources of moist static energy 
(or moist entropy) to the column.  The two most important such processes are 
surface turbulent fluxes and radiative cooling.  

Right or wrong, the requirement for moist static energy sources to be involved
in any linear instability is an interesting
feature of first baroclinic mode quasi-equilibrium models.
In extratropical atmospheric dynamics it has proved extremely useful to separate dry adiabatic
dynamical mechanisms (e.g., Hoskins et al. 1985)\nocite{hosminrob85} from
those in which "diabatic" processes (defined for a dry working fluid with
phase changes of water considered external), which break the conservation of potential 
temperature  and potential vorticity, are fundamentally involved.  It is clear
that dry adiabatic dynamics are inadequate to describe many important aspects
of the tropical atmospheric circulation and its variability, but the relative importance
of moist adiabatic dynamics --- as opposed to dynamics in which moist diabatic processes 
(those which break the conservation of moist static energy and moist entropy) are
critical --- remains unresolved.  The analogy to extratropical dynamics, and the
overall centrality of quasi-conserved variables in all of physics, suggests that
it is fruitful to ascertain the relative importance of moist adiabatic and diabatic
processes to intraseasonal variability.  This is a separate and more fundamental
question than that regarding the validity of first baroclinic mode QE models.  
Nonetheless those models provide a useful starting point for discussion since 
they make a clear prediction on the relevance of diabatic processes, 
as well as being both relatively tractable and based on principles
that are at root physically reasonable (convection acts to eliminate instability) 
even though some of their simplifying assumptions may be too strong 
for some purposes.

In the models of Emanuel (1987) and Neelin et al. (1987), 
Kelvin waves are destabilized by
the interaction of a convectively coupled wave with 
surface flux perturbations induced by the wave's surface wind 
perturbations.  This interaction was
called "wind-evaporation feedback" by Neelin et al. (1987) and "wind-induced
surface heat exchange (WISHE)" by Emanuel (1987).  For an
eastward moving Kelvin wave in a westward mean flow, a positive surface
wind speed anomaly occurs a quarter wavelength ahead of the location where 
the positive precipitation and vertical velocity anomalies would be in the 
absence of surface flux anomalies, but in phase with the
temperature anomaly.  Under the strict quasi-equilibrium assumption, the convection
responds immediately to surface flux anomalies, so the surface flux anomaly 
causes the heating anomaly to shift eastward, putting it partly in phase 
with the temperature anomaly and destabilizing the wave.

Key features of this theory for intraseasonal variability are that 
the waves must occur in an easterly mean surface flow, that the winds 
under the convective phase of the disturbance are easterly, and that the 
intraseasonal disturbances are Kelvin waves.  All of these features 
have been shown to be inconsistent
with observations.  It was immediately recognized  
that the strongest MJO events occurred in regions of mean westerlies\cite{wang_88,emanuel_88}.
It was then shown 
that the active phases, featuring enhanced 
precipitation, occurred in surface westerlies (e.g., Kiladis et al. 1994; Zhang and McPhaden 2000)
\nocite{kiladis_et_al_94,zhang_mcphaden_00}.  
Wheeler and Kiladis (1999)\nocite{whekil99} then showed that "convectively coupled" Kelvin
waves do exist, but that their spectral signatures are quite distinct from that
of the MJO, indicating that the two are different phenomena.  These observations
showed that the models of Emanuel (1987) and Neelin et al. (1987) are, in their
specifics, incorrect as explanations of the MJO.  

The observations are not, however, inconsistent with the general notion that
surface flux anomalies may be important to the dynamics of the MJO, but only
with the specific linear models proposed by Emanuel (1987) and Neelin et al. (1987).
If the disturbance is something other than a linear Kelvin wave, the 
requirements for mean easterly flow and net easterly flow in regions of 
active convection no longer apply.  Some studies with nonlinear models have 
identified such "nonlinear WISHE" as being important in simulated MJO-like 
disturbances \citep{ray01,malsob04}.

In the two decades since the publication of Emanuel (1987) and Neelin et al. (1987),
much more work has been done with idealized moist models which aim to explain
either the MJO, other aspects of tropical intraseasonal variability (such as the northward-propagating
mode found in northern hemisphere summer, discussed further below), or other parts of the 
convectively coupled wave spectrum.  In the case of the MJO in particular, none of these 
has been broadly accepted as providing a satisfactory explanation of the essential mechanisms
\cite{zhang_05,wang_05}.


\subsection{Northern summer intraseasonal variability}

In northern summer, intraseasonal variability modulates the Asian and
western Pacific monsoons. Spectra of atmospheric variability exhibit two significant peaks in the intraseasonal range: one at 10-20 days and one at 30-60 days \citep{Go05}. The 10-20-day mode is characterized by convective disturbances which propagate from the western Pacific warm pool and the maritime continent towards the northern Bay of Bengal and South Asia. These disturbances have been associated with equatorial Rossby waves deviated northward by the mean monsoon flow \citep{CG04}.   The 30-60-day mode is characterized by the northward propagation of approximately zonally-oriented rain bands  from $5\dg S$ to $25\dg N$  \citep{W&06}. This northward propagation is sometimes accompanied by eastward propagation  \citep{WR90, LW02}. Nevertheless, the northward propagating mode appears to be an independent regional mode of variability, rather than simply a local response in the South Asian region to the eastward-propagating disturbances \citep{JL05}, though this is
still controversial in some quarters (e.g., Sperber and Annamalai 2008).
  We focus here on this northward-propagating
mode, assuming that the eastward-propagating mode is essentially similar
to the southern summer MJO.  

Given the nearly zonal orientation of the rain bands and their nearly meridional direction
of propagation, a number of studies have assumed that longitudinal
structure is inessential to the dynamics of this mode, and modeled it axisymmetrically
 \citep{WC80,GS84,GS90,NSG92,SGW93,JLW04,DW05,BS07a,BS07b}.  
These studies have obtained linearly unstable northward propagating modes which
resemble the observed one to varying degrees.  In earlier studies, land-atmosphere interaction was proposed as crucial to the northward propagation \citep{WC80,W83}.  However, northward propagating modes were also later obtained in aquaplanet simulations 
\citep{GS84,NSG92}.  The northward propagation has been attributed in several recent studies to dynamical mechanisms that involve low-level convergence north of the propagating rainband.  This convergence is caused via 
Ekman pumping under a maximum of barotropic vorticity  which itself leads the maximum convection \citep{JLW04,Go05,BS07a,BS07b}. The mechanisms explaining the generation of this barotropic vorticity maximum are still debated \citep{JLW04,DW05,BS06,BS07b}. 

The question of what destabilizes the mode is distinct from that of what causes its
propagation.  In the model of \cite{BS07a,BS07b}, interactive surface fluxes were found to be
important to the instability of the northward propagating mode.  They used the quasi-equilibrium 
model developed by \cite{sobel_neelin_06}, which has a barotropic mode and prognostic boundary
layer in addition to a first baroclinic mode in the free troposphere.  Because of this more
complex vertical structure, the set of
possible dynamical mechanisms in this model is broader than that in the pure first baroclinic mode QE models.  It is possible for linear instability to occur in this model without surface flux feedbacks.
Nonetheless, Bellon and Sobel (2008a,b) found that WISHE
is critical to the linear instability of the northward-propagating
mode in the parameter regime which appears most justified based on observations.  
As usual with idealized models, one can easily challenge various details of this model (which has some similarities to earlier ones (e.g. Jiang et al. 2004) as well as some differences).  The results of Bellon and
Sobel (2008a,b) just show that it is possible to construct a plausible model of the 
northward-propagating mode of intraseasonal variability --- one based on physics that
is within the broad envelope of what is commonly found in idealized models of tropical
atmospheric dynamics, and also broadly consistent with observations  --- 
in which surface flux feedbacks are essential.  

\subsection{The near-equivalence of surface fluxes and radiation in quasi-equilibrium}
\label{sec:surface_and_radiative_fluxes}

The primary radiative effects of the high clouds associated with deep convection
are a cooling of the surface due to reflection and absorption of shortwave radiation
and a warming of the atmosphere due to the greenhouse effect in the longwave
and the absorption of shortwave.  To the extent that these effects have similar
magnitudes, so that they cancel at the top of the atmosphere, they lead to a 
cooling of the ocean and equal warming of the atmosphere. 
This is equivalent to a surface flux, as far as the vertically integrated energy
budget of the atmosphere is concerned.  QE theory provides a useful
context in which to frame this equivalence.

If the vertical structure of the atmospheric flow is assumed fixed
(for example, a first baroclinic mode), and if we assume steady state
and neglect horizontal advection, the budget of moist static energy
requires that  the large-scale vertical motion, or net vertical mass flux, is 
proportional to the net convergence of the vertical flux of moist static energy 
into the tropospheric column (e.g., Neelin and Held 1987;  Raymond 2000;  Neelin 1997;
Neelin 2007;  Sobel 2007)  \nocite{neehel87,ray00,neelin_97,neelin_07,sobel_07}.  
The latter is the sum 
of the net turbulent latent and sensible surface heat fluxes plus the vertically
integrated radiative heating of the troposphere (or minus the radiative cooling).  

The proportionality factor which relates the energy flux to the mass flux is known 
as the gross moist stability (GMS), following \cite{neehel87}.   There is no very good 
theory for the value of the GMS, though some observational estimates have been made 
\citep{yuet98,back_bretherton_06}.  
The first baroclinic mode assumption is restrictive, perhaps even qualitatively misleading in
some circumstances (e.g., Sobel 2007), but no better idea of comparable simplicity has
yet appeared.  In general, the GMS need not be a constant or a simple
function of the temperature and humidity profiles alone (as in first baroclinic mode
QE theory), because it is quite sensitive to the
vertical profile of the divergent circulation \citep{sobel_07}.  Since the latter 
can vary dynamically on a range of space and time scales, the GMS
can as well.  In simulations in a GCM with simplified physics~\cite{fri07a} the GMS is
strongly influenced by properties of the convective 
parameterization~\cite{fri07b}.

For our immediate purpose,
what matters most is that GMS be positive on intraseasonal time scales, 
so that increases in net 
vertical energy flux convergence into the column lead (with a lag that is either negligible
or at least short by comparison to the intraseasonal timescale;  storage on timescales
of a few days does not significantly complicate the argument) to increases in
vertical mass flux, which in turn imply increases in deep convection. 
This is a weaker constraint than usually assumed in QE theory, though the difference 
is one of degree rather than kind.  Even the positivity of the gross moist stability
is questionable in observations, particularly in the eastern Pacific ITCZ
(Back and Bretherton 2006), but it appears to be a reasonable assumption in the
Indian and western Pacific regions.

We assume that the difference in the cloud field
between convectively active and suppressed precipitation regimes consists predominantly
of the presence vs. absence of high clouds.  Satellite observations have
shown that, in the mean these clouds produce perturbations
in the net radiative energy flux at the top of the atmosphere which are small
compared to their largely cancelling shortwave and longwave components
~\citep{ramanathan_et_al_89,harrison_et_al_90,hartmann_et_al_01}.  Lin
and Mapes (2004) found that this cancellation is less close on intraseasonal
time scales, with MJO-related shortwave anomalies being larger than
longwave ones by as much as 30\%.  This is a significant difference, but still
the cancellation substantially exceeds the remainder.  The
implication is that any anomalous radiative heating of the atmosphere due to these clouds,
whether occurring in the longwave or shortwave bands, is approximately
compensated by anomalous radiative cooling  
of the ocean.  In the vertically integrated moist static
energy budget, cloud-radiative
heating anomalies due to deep convection are essentially similar to
convectively induced perturbations to turbulent surface heat fluxes, as
both amount to a net transfer of energy from ocean to atmosphere
in a convectively active phase.  When convection is active, there is a
net decrease of radiative energy flux into the ocean, accompanied by
a significantly smaller change in the top-of-atmosphere balance.  Thus we use the phrases
"surface fluxes" or "surface flux feedbacks" to include radiative cooling feedbacks.

\subsection{Ocean coupling}

A substantial body of work over the last decade or so argues 
that intraseasonal SST variability is not only
driven by the atmosphere, through intraseasonal variations in surface
energy fluxes, but that SST variability also influences the atmosphere 
through the influence of SST anomalies on column stability and deep convection.
To the extent that these feedbacks are significant, intraseasonal variability
is coupled.  Most GCM studies addressing this in the context of the MJO
have shown some enhancement of the simulated variability
in experiments with atmospheric models coupled to a mixed-layer
ocean models, as compared to models with fixed SST\citep{Waliser-Lau-Kim-1999:influence,kemball-cook_et_al_02,zheng_et_al_04,fu_et_al_07}, although at
least one study found no enhancement~\citep{Hendon-2000:impact} and
and others found small enhancements (e.g., Maloney and Sobel 2004) or
mixed results, with differences in the mean climate between coupled and
uncoupled runs complicating the interpretation~\citep{inness_slingo_03}.
This evidence suggests that the MJO
is enhanced by coupling, but is not fundamentally dependent on coupling
for its existence.  In virtually all models tested in this way, a simulated
MJO is present to some degree without coupling.

Observations suggest that
coupling has a qualitatively similar impact on intraseasonal variability 
of the Asian monsoon in northern hemisphere summer, including northward-propagating
rainbands and SST variability in the Arabian sea and Bay of Bengal~\citep{Vecchi-Harrison-2002:monsoon,W&06,roxy_tanimoto_07}.  
In GCM studies, ocean coupling enhances northward-propagating intraseasonal
variability in the Indian Ocean to varying degrees 
(e.g. Zheng et al. 2004, Seo et al. 2007, Fu et al. 2007,
Fu and Wang 2004, Kemball-Cook et al. 2002)\nocite{kemball-cook_et_al_02,seo_et_al_07,zheng_et_al_04}.  
One recent study with an idealized axisymmetric model
suggests that the SST variability is largely passive, being forced by
the atmosphere but having only a modest impact on the atmospheric mode~\cite{BSV08}.

The question of the importance of ocean coupling is related to but not the same as
that of the importance of surface fluxes to the dynamics of intraseasonal variability.  If ocean 
coupling is important, surface fluxes must be involved, since 
only through those fluxes can the ocean influence the atmosphere.  
The converse is not true:  an important role for surface fluxes does not necessarily
imply that coupling is important.  Surface flux feedbacks can operate in models which 
assume fixed SST.  Such models do not satisfy a surface energy budget, but their surface
fluxes can still vary interactively and influence the atmosphere.  Coupling can either
amplify or damp intraseasonal variability, depending on the phasing of the SST 
anomalies relative to anomalies in atmospheric variables.
For example, Shinoda et al. (1998)\nocite{Shinoda-Hendon-1998:mixed} 
found that observed
SST anomalies slightly reduced the amplitude of MJO-related surface latent 
heat fluxes compared to what they would have been for fixed SST.

Our interest here is in the role of surface fluxes in the dynamics of atmospheric
intraseasonal variability.  Ocean coupling, while also arguably important, is secondary
in this discussion.  However, as discussed next, the nature of the underlying surface
is important to the extent that it must have sufficiently large heat capacity to allow
substantial fluctuations in the net surface enthalpy flux on the intraseasonal time scale.

\subsection{Single column dynamics}
\subsubsection{Single column dynamics inferred from observations}
\label{sec:scm_obs}
The MJO is commonly defined as having large spatial scales.
However, plots of intraseasonal variance in quantities related to deep
convection also show relatively small-scale features as described in section 3.
These smaller-scale features appear to be related to the nature
of the underlying surface, and thus dynamically to be a result of the local 
interaction of that surface with the atmosphere.  A simple framework
within which to grasp
these local interactions may be the idealized dynamics
of a single column, consisting of the atmosphere and ocean in
a relatively small horizontal area.  

The single-column view is taken in the observational study of 
\cite{Waliser-1996:formation}, who
showed in a composite analysis that an oceanic ``hot spot'', defined as
a region of at least $1\times10^6$ km$^2$ in which the sea surface temperature (SST) 
exceeds 29.75$^\circ$ for at least a month, 
typically appears after period of calm surface winds and clear skies.
Anomalously strong surface winds and
enhanced high cloudiness develop after the time of peak SST.  
Associated surface latent heat flux and shortwave radiative
flux anomalies lead to the decay of the hot spot. 

The conceptual model resulting from Waliser's study, as well as similar ones
articulated by subsequent studies~\citep{fasullo_webster_99,stephens_et_al_04},
describe coupled oscillations occurring in a single column. They leave open to what extent
the oscillation can be self-contained in a single region, 
as opposed to being fundamentally driven by the passage of large-scale
disturbances. 
Observations suggest that the latter is a better description of the MJO (e.g., 
Hendon and Glick 1997; Zhang and Hendon 1997), 
\nocite{Hendon-Glick-1997:intraseasonal,Zhang-Hendon-1997:propagating} 
as well as the northward-propagating Asian monsoon mode, since they have
large-scale spatiotemporal structure and propagation within which
regional-scale features are embedded. Nonetheless, the single-column view is useful
for understanding some aspects of the controls on deep convection, and may be
particularly relevant to understanding the smaller-scale regional features shown 
below.

\subsubsection{A simple coupled single-column quasi-equilibrium model}

Sobel and Gildor (2003, SG03) \nocite{sobel_gildor_03} presented an explicit single column 
dynamical model of a simple atmosphere coupled to a slab mixed-layer ocean of 
constant depth.  This model was designed to capture the behavior described by the 
observational studies described in section \ref{sec:scm_obs} above.  Their model 
incorporates the standard assumptions of first baroclinic mode QE theory, as expressed 
in the  ``quasi-equilibrium tropical circulation model'' (QTCM)
formulation \cite{neezen00,zeng_et_al_00}, plus a few additional assumptions. 
The most important additional
assumption is that the local temperature
profile --- that is, not only the vertical structure of the temperature
field, but also its value --- is fixed.  In this  "weak temperature
gradient" (WTG) approximation (e.g., Held and
Hoskins 1985;  Neelin and Held 1987;  Mapes and Houze 1995;
Zeng and Neelin 1999;  Sobel and Bretherton 2000;
Sobel et al. 2001;  Majda and Klein 2003) \nocite{helhos85,maphou95,majkle03,sobel_bretherton_00,sobet01,zeng_neelin_99}
large-scale vertical motion is assumed to occur
as needed in order for adiabatic cooling to balance diabatic heating, and 
the local temperature profile is held close to that of surrounding regions by a
process which is essentially geostrophic adjustment with a small Coriolis
parameter \cite{bretherton_smolarkiewicz_89}.  This parameterization
of large-scale dynamics allows the precipitation 
to vary strongly in response to variations in SST, surface turbulent fluxes, and 
radiative cooling.  In the absence of a large-scale circulation --- for
example, in radiative-convective equilibrium --- large precipitation variations 
cannot occur, because any changes
in convective heating have to be balanced by changes in radiative cooling,
which cannot become too large.   By employing the WTG approximation rather
than an assumption of no large-scale circulation (as in, for example, Hu and
Randall 1994, 1995)\nocite{hu_randall_94,hu_randall_95}, 
the SG03 model allows physically plausible, if highly
parameterized, interactions 
between convection, radiation 
and large-scale dynamics to occur in a single column.

SG03 also assumed that
deep convective clouds induced radiative perturbations which reduced
the longwave cooling of the atmosphere and shortwave warming
of the ocean surface in equal measure, so that the cloud-radiative perturbations 
play a role very similar to that of surface flux perturbations, as discussed above.
SG03 very crudely represented the effect of surface
flux feedbacks by assuming them to be local and lumping them together
with radiative feedbacks, which in turn were parameterized as
proportional to precipitation. They 
did this by increasing the proportionality coefficient, {\em r}, relating
radiative anomalies to precipitation anomalies compared to that
estimated from observations~\citep{bresob02,lin_mapes_04},
arguing that the parameterized flux anomalies represented those in
both radiative and wind-induced turbulent surface fluxes.
Recent work suggests such increases are justifiable.
Araligidad and Maloney (2008) found 
that intraseasonal latent heat flux anomalies alone are about 20\% of
precipitation anomalies in the west Pacific warm pool (e.g., 
\nocite{araligidad_maloney_08}, which appears at broadly consistent
with the wind speed-precipitation relationship found on daily time scales
by Back and Bretherton (2005)\nocite{back_bretherton_05}.

With a proportionality coefficient of around 0.25 or greater ---
corresponding to a net 25 W m$^{-2}$ transfer of energy from ocean
to atmosphere for each 100 W m$^{-2}$ of column-integrated latent
heating --- and other parameters set at typical control values,
 the model of SG03 is linearly unstable to free
oscillations which qualitatively resemble those found in the
observational studies (Waliser 1996; Fasullo and Webster 1999; 
Stephens et al. 2004).  The growth rate of the oscillations
is sensitive to several parameters in the model, including the
surface enthalpy flux feedback parameter $r$, the
mixed layer depth, the time scale for convective adjustment,
and the GMS (which is assumed constant), 
but the period is robustly in the intraseasonal range.


SG03 argued that their model
on its own was not adequate to represent the MJO, as its
single-column structure makes it incapable of capturing the
MJO's horizontal structure and propagation.  They argued
instead that their model might better depict how a small
 horizontal area responds to the passage of
an MJO disturbance, with that disturbance viewed as 
an external forcing.  With $r$ small enough
to render the model stable --- such as is appropriate if it
represents radiative feedbacks alone --- the model solution is the 
forced response of a damped oscillator.   SG03 imposed the
forcing through the atmospheric temperature field, which
they took to have a sinusoidal variation with intraseasonal
period.  Maloney and Sobel (2004) instead imposed an intraseasonally
fluctuating surface wind speed forcing in their application of the SG03 model,
making it appropriate to set $r\sim 0.1$, representing radiation only.  

\begin{figure}
\noindent
\includegraphics[height=5cm]{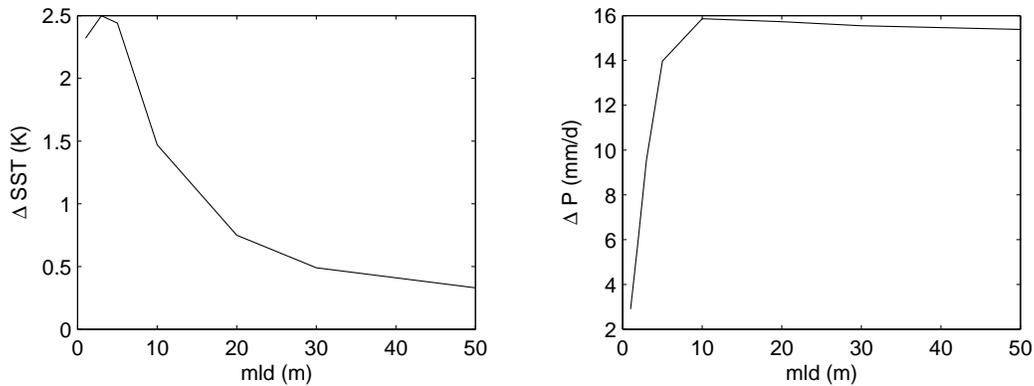}
\caption{Precipitation amplitude vs. mixed layer depth, SG03 model,
from Maloney and Sobel (2004).}
\label{fig:ms04_sg03}
\end{figure}

A parameter of particular interest in this model is the mixed layer
depth.  The amplitude of the model oscillations in precipitation 
as a function of mixed layer depth is 
presented in Figure~\ref{fig:ms04_sg03}, 
reproduced from Maloney and Sobel (2004).  
In this curve, the precipitation amplitude has a maximum at a particular
value of the mixed layer depth, around 10-20 meters.
It falls off slowly as mixed layer depth increases past the 
maximum, and rapidly as the mixed layer
depth approaches zero.
A similar (if weaker) amplitude maximum was found in the 
GCM results of Maloney and Sobel (2004) and is supported
by a recent analysis of spatial and seasonal variability in
the amplitude of intraseasonal variability compared to that
in mixed layer depth~\citep{bellenger_duvel_07}.  

The amplitude decrease for mixed layers deeper than that
at which the maximum occurs indicates that in this model, ocean
coupling can (modestly) enhance intraseasonal variability, since infinite mixed layer depth
corresponds to fixed SST.  This decrease was also found in the
GCM study of Watterson (2002)\nocite{watterson_02}, and is implied in those
studies which find stronger intraseasonal variability in coupled models than
in atmospheric models over fixed SST.

The vanishing of the response
as mixed layer depth goes to zero reflects the fact that surface
enthalpy fluxes are critical to the 
oscillations in this model.  As the mixed layer depth
approaches zero, the net surface enthalpy flux must also 
vanish.  This kills the oscillation because the gross moist stability is
positive, requiring net moist static energy input into the column
(which is equivalent to net surface enthalpy flux under our
assumptions) in order to generate circulation anomalies.

A mixed layer of zero depth, or "swamp", with zero heat capacity 
but an infinite moisture supply, may be thought of as
a crude representation of a land surface in a tropical region during
its monsoon season, although it is not a good representation of 
land surface processes in general.  When soil becomes subsaturated,
variations in the Bowen ratio (ratio of sensible to latent heat flux) can result
in "soil moisture memory" by which the land-atmosphere
interactions have intrinsic time scales of up to several months.
This effect appears most important in semi-arid regions~\citep{koster_suarez_01}.  
In active monsoon regions,
soil moisture memory is less important, and we assume that modeling the land surface
dynamics by a swamp, with zero heat capacity but infinite available
moisture, is adequate for purposes of understanding the qualitative
dynamics of the coupled system.  Treating each horizontal location as
represented by an independent single column model under WTG and
SQE (SG03), we arrive at the prediction that {\it the amplitude of intraseasonal
variability in deep convection can vary locally depending on surface type,
and should be small over land and larger over ocean.}

\section{Observations}
\label{sec:obs}
\subsection{Intraseasonal variance maps}

\subsubsection{Results}

\begin{figure}
\noindent
\includegraphics[height=5cm]{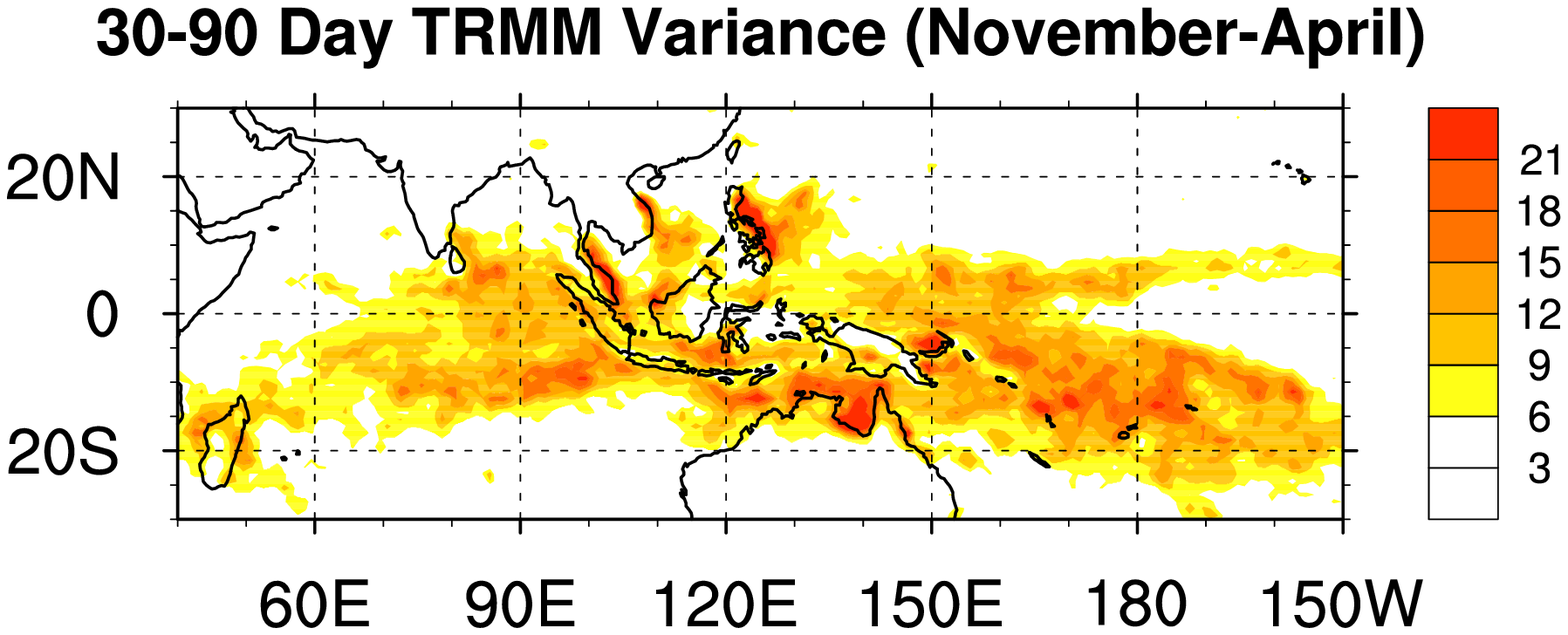}
\includegraphics[height=5cm]{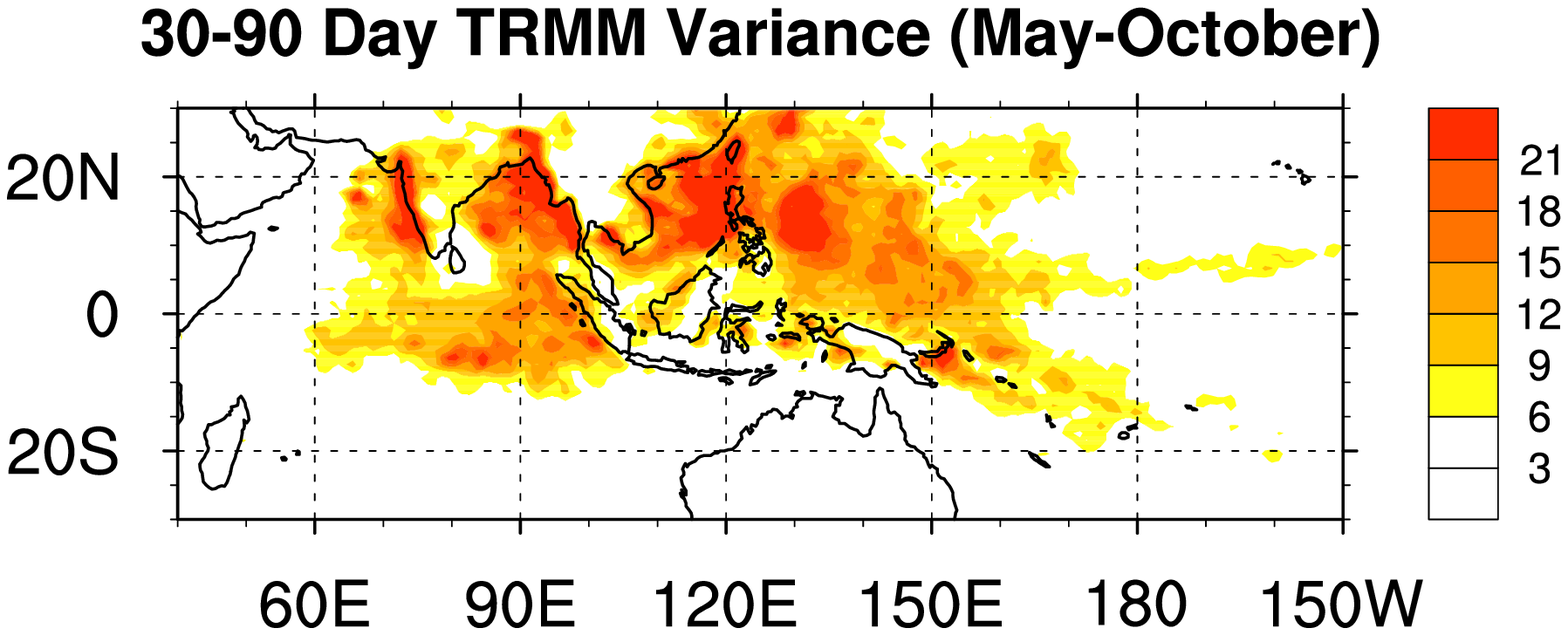}
\caption{Intraseasonal variations in rainfall for a) November-April
and b) May-October ($mm^2~d^{-2}$).}
\label{fig:rain_variance}
\end{figure}

\begin{figure}
\noindent
\includegraphics[height=5cm]{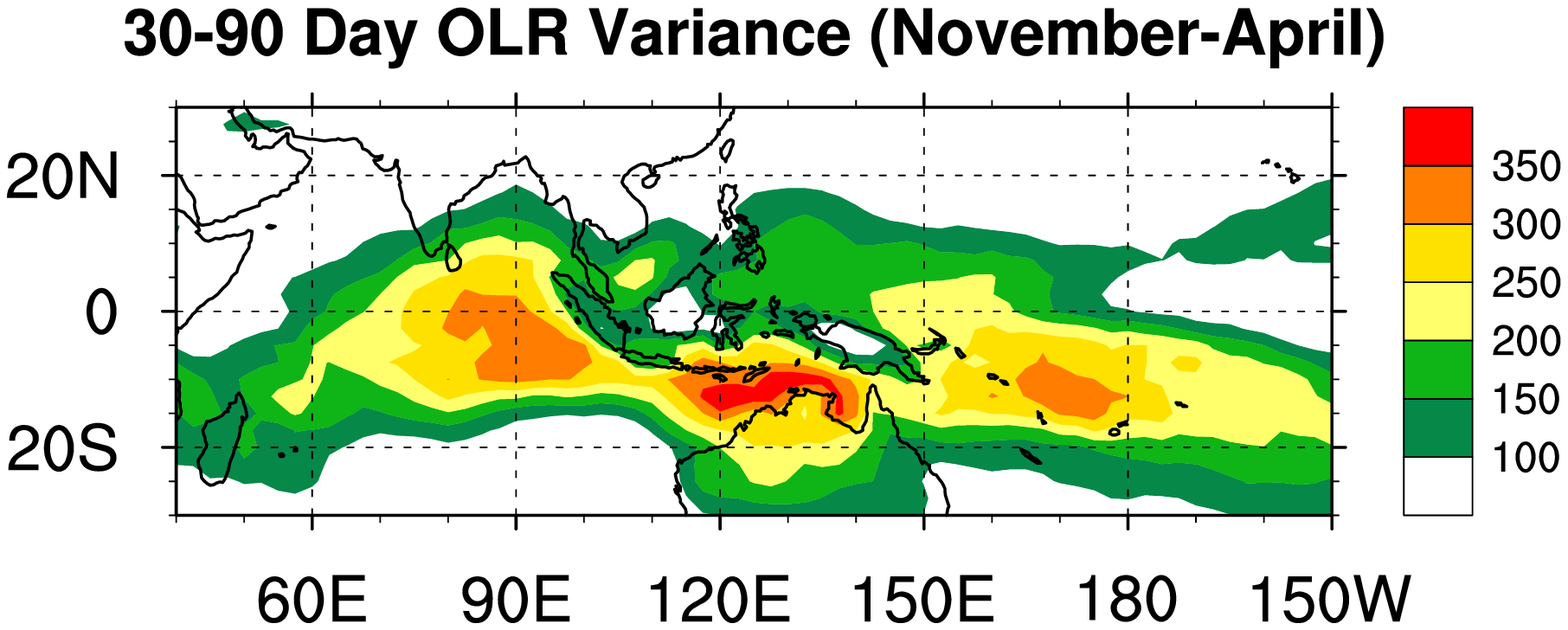}
\includegraphics[height=5cm]{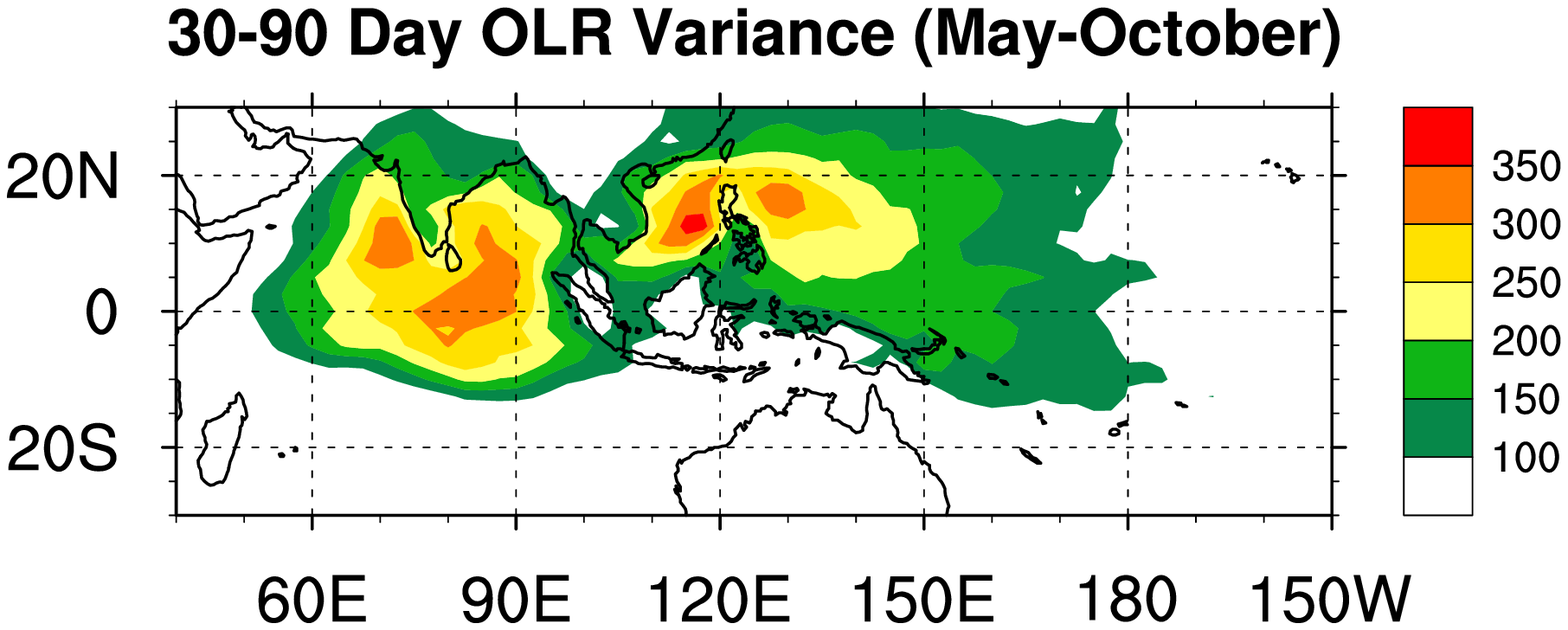}
\caption{Intraseasonal variations in OLR for a) November-April
and b) May-October ($W^2~m^{-4}$).}
\label{fig:olr_variance}
\end{figure}

Figures~\ref{fig:rain_variance} and~\ref{fig:olr_variance} show maps of
30-90 day variance in precipitation from the TRMM 3B42 precipitation 
data set and outgoing longwave radiation (OLR) from the NOAA interpolated OLR 
data set, respectively, for the months November-April and May-October.  
 Similar maps are shown in previous studies (e.g., Weickmann et al. 1985;  Zhang and
 Hendon 1997;  Vincent et al. 1998;  Fasullo and Webster 1999;  Sperber 2004).  
  \nocite{weickmann_et_al_85,vincent_et_al_98,sperber_04}
Daily-averaged TRMM precipitation data during 1998-2005
averaged to a $1^\circ \times 1^\circ$ grid are used. The TRMM 3B42 product we use here incorporates
several satellite measurements, including the TMI and TRMM precipitation radar to
calibrate infrared measurements from geostationary satellites~\citep{adler_et_al_00}.
The daily-averaged
NOAA interpolated OLR product is used during 1979-2005 on a $2.5^\circ \times 2.5^\circ$ grid~\cite{liebmann_smith_96}. Unless otherwise stated, intraseasonal bandpass filtering is
conducted using two 60-point non-recursive digital filters with half-power points at 30 and
90 days.

During southern hemisphere summer, intraseasonal 
variability is dominated by the canonical MJO, which has very large horizontal scales (Figs.~\ref{fig:rain_variance}a and~\ref{fig:olr_variance}a).  
However, the variance maps also exhibit prominent
smaller-scale patterns.  These small-scale patterns
consist primarily of enhanced intraseasonal variance over the oceans and reduced variance
over land.  Particularly striking is the land-sea contrast in the Maritime Continent
region.  This land-sea variance contrast is also prominent in northern summer (~\ref{fig:rain_variance}b
and~\ref{fig:olr_variance}b), and
is evident down to the smallest
scales resolved by the data.

\begin{figure}
\noindent
\includegraphics[height=5cm]{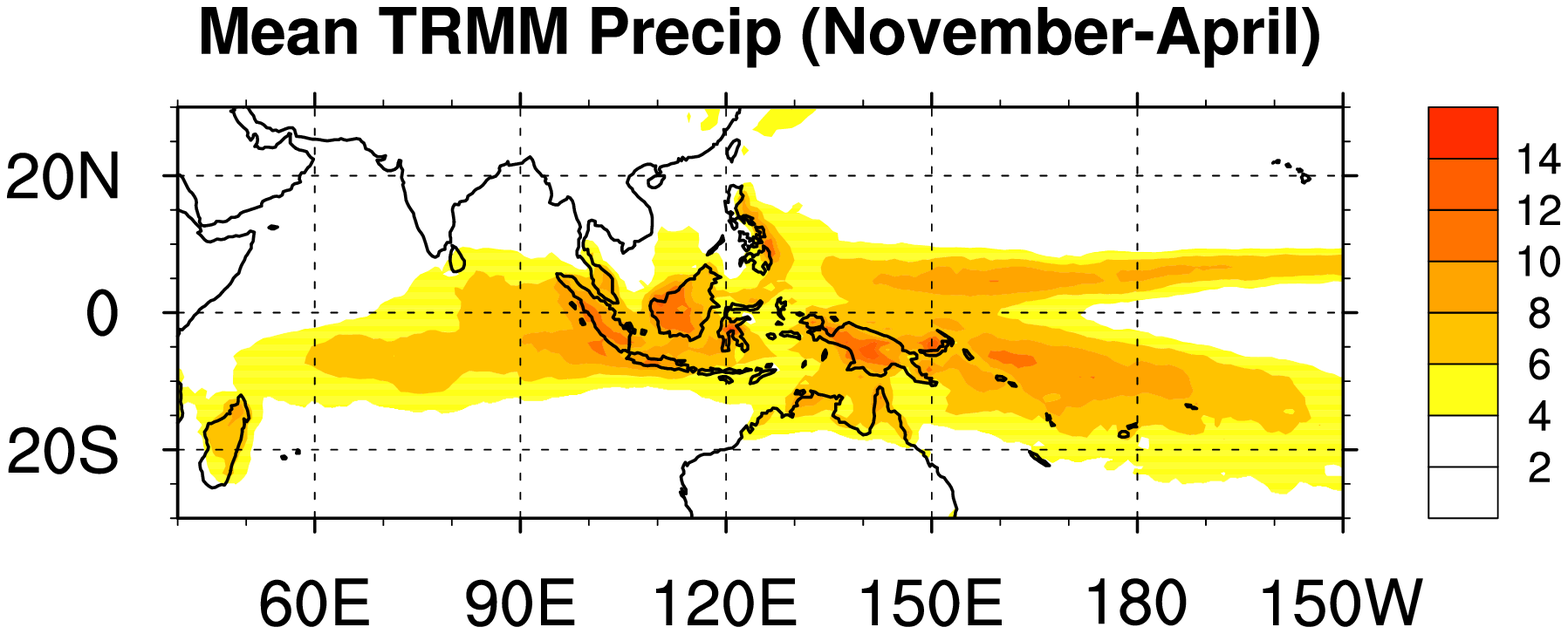}
\includegraphics[height=5cm]{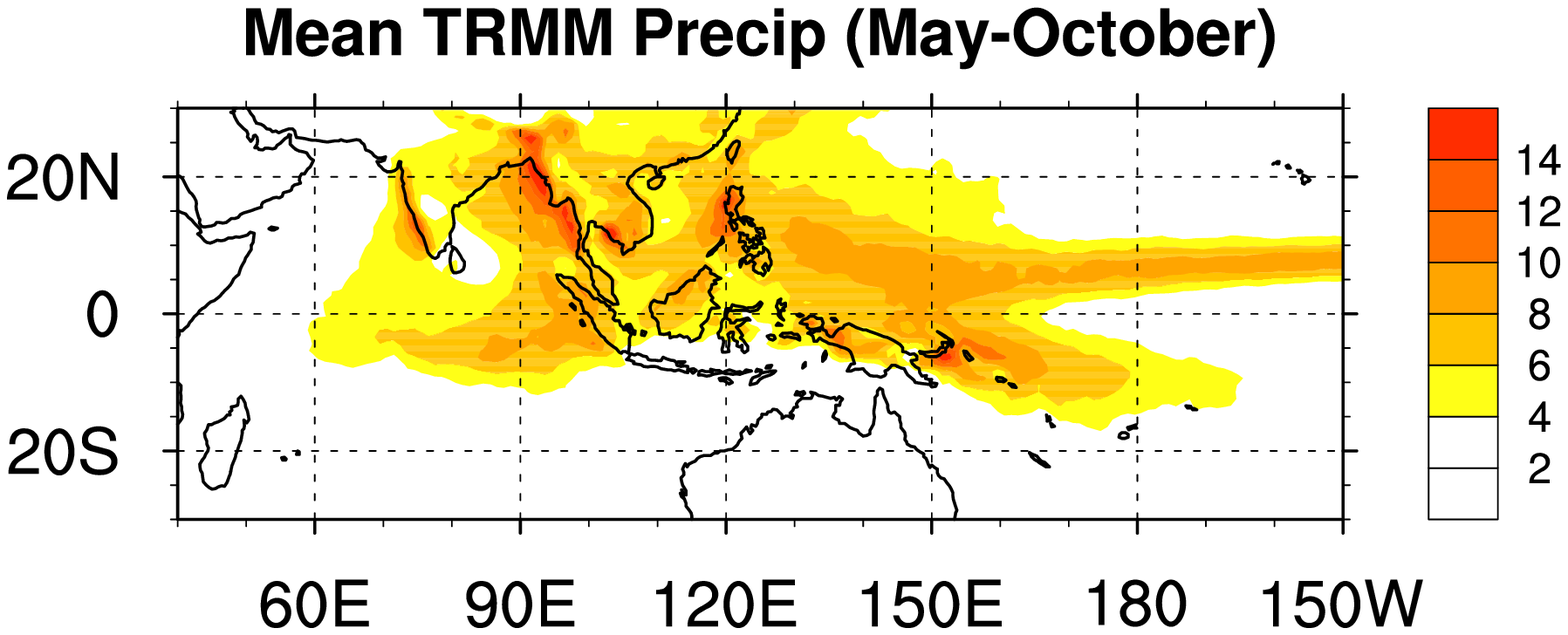}
\caption{Climatological rainfall for a) November-April
and b) May-October ($mm~d^{-1}$).}
\label{fig:rain_climo}
\end{figure}

\begin{figure}
\noindent
\includegraphics[height=5cm]{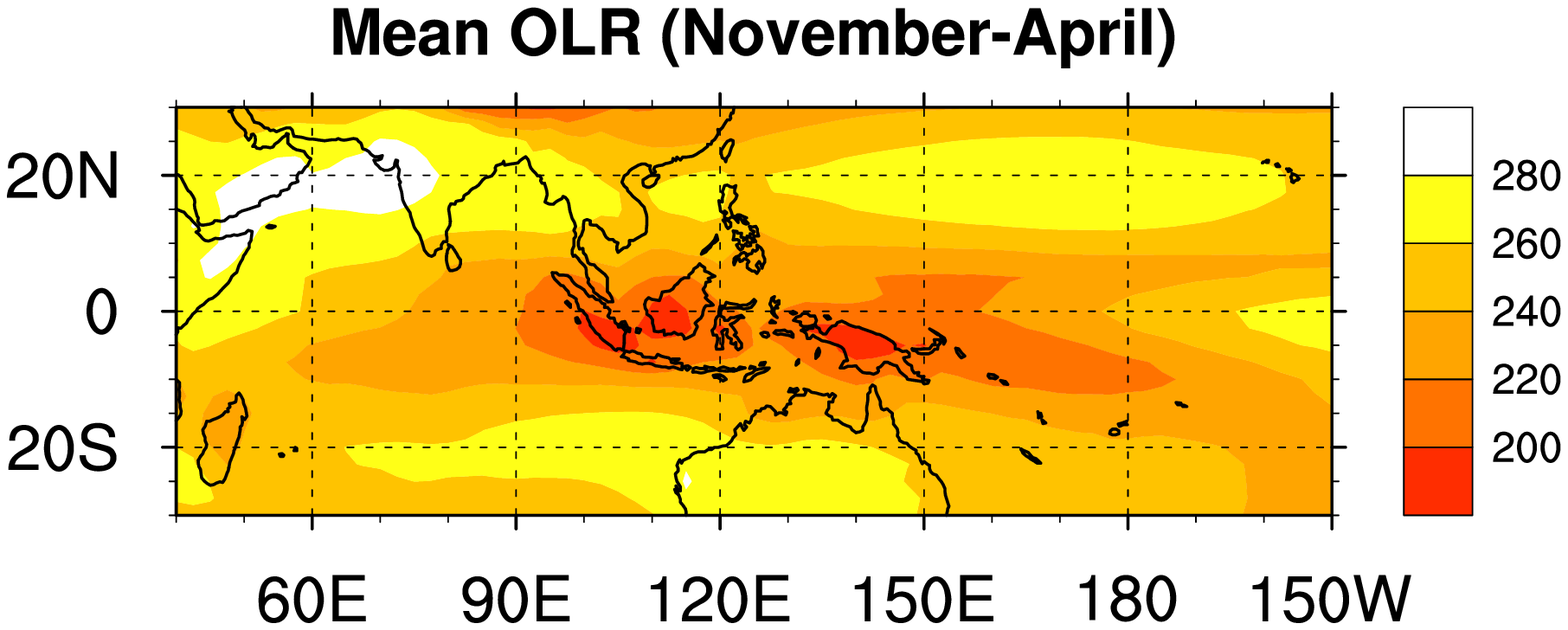}
\includegraphics[height=5cm]{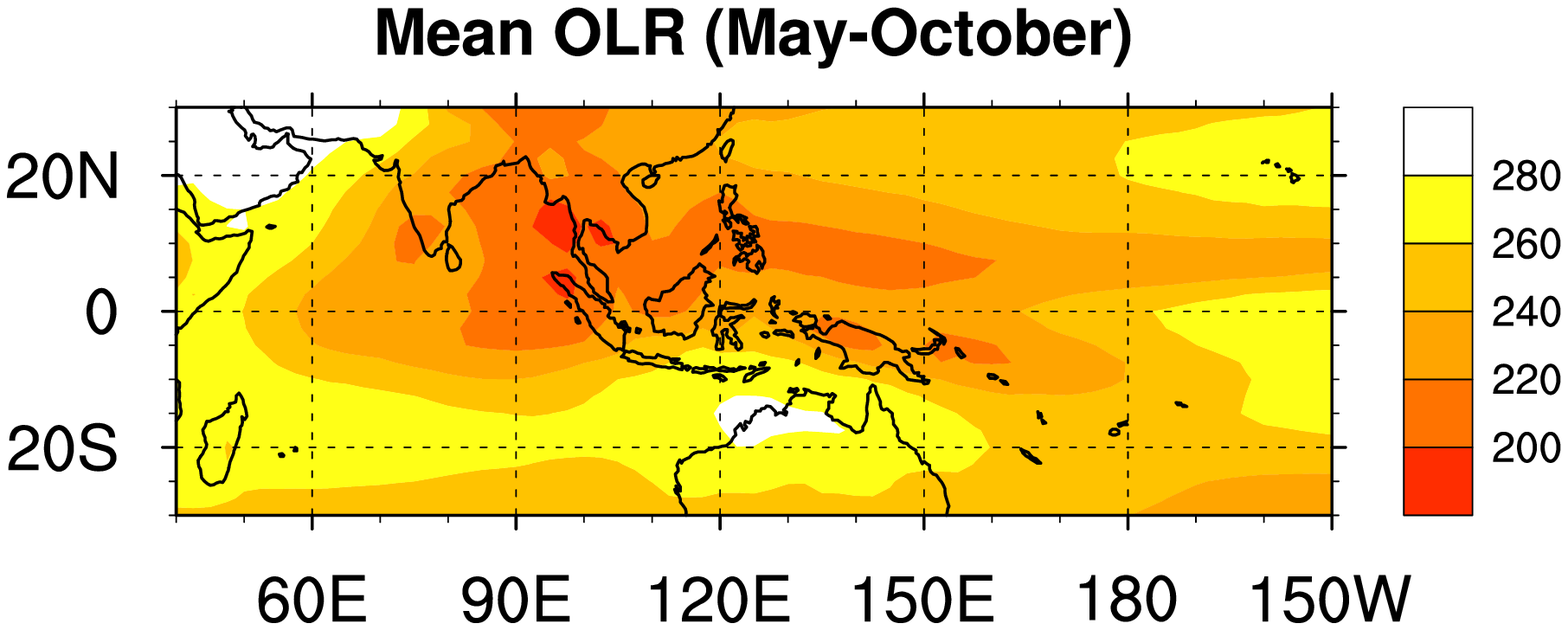}
\caption{Climatological OLR for a) November-April
and b) May-October ($W~m^{-2}$).}
\label{fig:olr_climo}
\end{figure}

Figures~\ref{fig:rain_climo} and~\ref{fig:olr_climo}
show the climatological mean precipitation and OLR, respectively for May-October and
November-April.  The patterns of the southern and northern hemisphere
monsoons are evident.  In May-October (\ref{fig:rain_climo}b and~\ref{fig:olr_climo}b), 
the climatological precipitation
resembles the intraseasonal variance in its horizontal structure, 
with maxima in rainfall over the oceans and minima 
 over land.  In November-April,
however, the same is not true (\ref{fig:rain_climo}a and~\ref{fig:olr_climo}a).  
Climatological convection maximizes
over the large islands of the maritime continent region, while intraseasonal
variance minimizes there.  This tendency is most striking when examining
the OLR product, although
neither the patterns of intraseasonal variance nor those of climatological precipitation
shown above is sensitive to the choice of data set.  Similar patterns are apparent, for example,
 in the CMAP \cite{Xie-Arkin-1997:global}
 precipitation data set (not shown).  

\subsubsection{A proposed explanation of observed variance patterns}

Despite the extreme simplicity of the SG03 model, the small-scale features in the 
observed patterns of intraseasonal precipitation variance are consistent
with it, and thus with the assumptions of convective quasi-equilibrium and WTG which
it incorporates.  This is true in
at least two important respects:
\begin{enumerate}
\item The fine-scale structure in precipitation variance suggests that, despite
the large-scale structure of the flow features associated with intraseasonal
variability, it may be appropriate to consider variations in convection in terms
of a local picture, which can be captured by a single-column model using WTG
[or perhaps also by other single-column parameterizations of large-scale dynamics (e.g.,
Bergman and Sardeshmukh 2004; Mapes 2004)]\nocite{mapes_04,bergman_sardeshmukh_04}.
\item The fact that intraseasonal precipitation variance maximizes over the
ocean and minimizes over the land is consistent with an important role for
interactive variations in the net surface enthalpy flux in generating the variance,
since such variations can have significant amplitude over ocean but not over
land.
\end{enumerate}

The variations in net enthalpy flux most likely have turbulent and radiative components,
corresponding to wind-evaporation and cloud-radiative feedbacks.  It is not possible
to determine which is more important on the basis of either the idealized
model of SG03 or the patterns of variance alone.  The observational analysis of Waliser (1996)
suggests that the two components may be of comparable magnitude, while Hendon and
Glick (1997) suggest that the relative importance of the two may vary with location.  

\subsubsection{Alternative explanations}

An alternative explanation for the patterns shown in Figs.~\ref{fig:rain_variance} and
~\ref{fig:olr_variance} is 
that the patterns are controlled by orographic effects rather than by land-sea contrasts
in surface enthalpy fluxes.
For example, in May-October, the patterns of intraseasonal precipitation variance 
(especially away from the maritime continent) resemble
the patterns of mean rainfall, which are certainly influenced by orography.  The
orographic influence most likely is largely due to the dynamical forcing of
upslope flow as monsoon winds impinge on mountain ranges, and is thus
dynamically distinct from the thermodynamic effects of land-sea contrasts.  Focusing
on the Indian and Southeast Asian regions, one maximum (in both variance and
mean rainfall) occurs over and just upstream of the Western Ghats, while another
occurs over the Bay of Bengal, upstream of the mountains on the Cambodian coast.
It might be argued that the variance minimum in between (fig. \ref{fig:rain_variance}b), 
over the Indian subcontinent,
owes its existence to the minimum in mean rainfall, and that the latter minimum owes its existence
to orography.  Southern India lies in the rain shadow of the Western Ghats, making it
drier than the regions upstream and downstream.  In general, where mean rainfall is smaller
variability will also be smaller.  A related argument follows
from the analysis of Hoyos and Webster (2007)\nocite{hoyos_webster_07}, 
who present evidence both that
much of the total precipitation falling in the Asian monsoon is associated with intraseasonal
events, and that the precipitation distribution in these events is modulated by orography.

While orography undoubtedly influences the rainfall patterns shown above, 
orographic effects alone
cannot explain all aspects of the intraseasonal variance maps shown in Figures~\ref{fig:rain_variance}
and ~\ref{fig:olr_variance}.
We contend that these patterns can be explained more generally by the
land-sea difference in heat capacity, 
which in turn suggests a role for surface fluxes.  This is particularly apparent
when we consider the November-April and May-October results together, and look for
the most general explanation for both.  Consider the maritime continent region in November-April.
Grossly speaking, 
intraseasonal variance maximizes over ocean and minimizes over land.
Mean rainfall does the opposite, particularly over the largest Indonesian islands (though
there is also more complex structure within individual islands which is surely influenced
by topography).  The large mountains on these
islands most likely play a role in inducing the mean precipitation
maxima (e.g., Qian 2008)\nocite{qian_08}.  
If the structure of the intraseasonal variance were determined
by the structure of the mean rainfall, we would expect to see variance maxima over these
large islands, coincident with the mean rainfall maxima, but instead these are regions
of relatively low variance.  The dominance of surface type over orography in the determination
of the intraseasonal variance patterns is also suggested by
the pattern over and around northern Australia, where intraseasonal variance also 
maximizes over ocean to a greater extent than mean rainfall does.  Northern Australia lacks
significant orography, so it seems almost certain that this difference is due to
land-sea contrast.  Even in May-October, a primary role for
land-sea difference in heat capacity rather than orography is suggested by the maximum variance
to the east of the Philippines, which lies neither over nor immediately 
upstream of any mountains.

Besides orography, 
another explanation might be that convection over land is dominated by the diurnal 
cycle, and this disrupts the variability at intraseasonal time scales
(e.g., Wang and Li 1994)\nocite{wang_li_94}. This can be interpreted 
as essentially the same explanation that we present in section 2e, but in 
different language. If we force
the SG03 single column model (for example) with insolation which varies
at diurnal frequencies, the response maximizes at a very small mixed layer depth (not shown). 
Over land, due to the small heat
capacity of the surface, the preferred frequency for coupled single-column oscillations is
much higher than over ocean. Thus, the system responds more strongly to diurnal forcing
and less strongly to intraseasonal forcing.

\subsection{Correlation between surface latent heat flux and precipitation}

Surface fluxes can be important to intraseasonal variability only if anomalous
surface fluxes are able to influence the occurrence or intensity of deep convection.
If this is the case, we might reasonably
expect surface fluxes and precipitation to covary in space and time.  The degree of
covariance has been assessed in a couple of recent studies.  Back and Bretherton (2004)
showed that there is a small but significant correlation on the daily time scale
between surface wind speed 
(which plays the dominant role in controlling the surface turbulent flux variations over
tropical oceans) and precipitation, with the correlation being stronger for regions of
high column water vapor content.

\begin{figure}
\noindent
\includegraphics[height=10cm,clip]{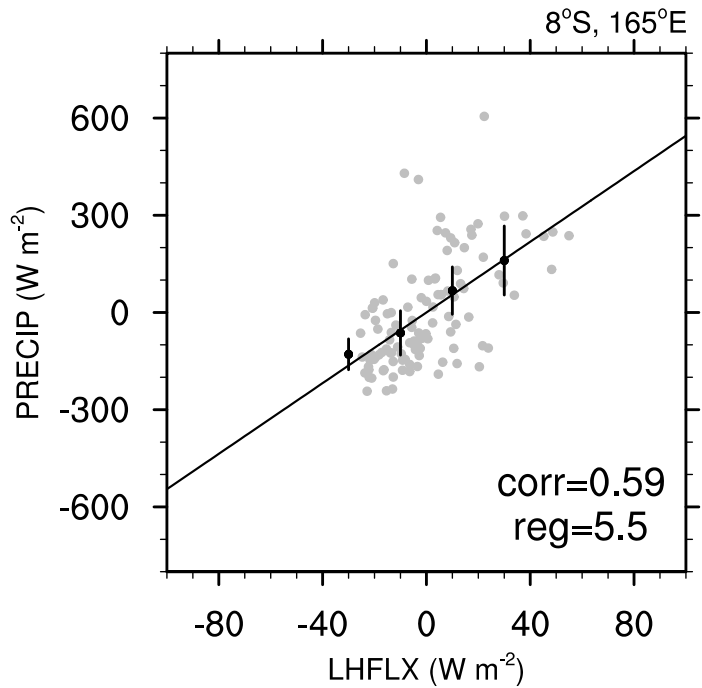}
\caption{Scatterplot of intraseasonal (15-90 day filtered)
TRMM 3B42 precipitation vs. TAO buoy latent heat flux at $8^\circ$S, $165^\circ$ E,
from Araligidad and Maloney (2008).  Regression and correlation coefficients are indicated on the plots, and the black points represent binned averages, with bars depicting the 90\% confidence limits about those averages.}
\label{fig:scatter_lh_precip}
\end{figure}

On the intraseasonal time scale, surface latent heat flux and precipitation (or
quantities related to it, such as OLR) have
been found to be locally correlated.  The peak correlation is typically found when
latent heat flux lags convection by a week or so, though that optimal lag varies slightly
from one study to the next (e.g., Hendon and Glick 1997,
Shinoda et al. 1998, Woolnough et al. 2000)
\nocite{shinoda_et_al_98,Woolnough-Slingo-Hoskins-2000:relationship}.
Araligidad and Maloney (2008) demonstrated a strong instantaneous correlation (~0.7)
between November-April 30-90 day QuikSCAT wind speed and TRMM precipitation
within the west Pacific regions of strong intraseasonal precipitation variance shown in Figures 2 and 
3.  Araligidad (2007) demonstrated similar strong correlations in the Indian Ocean during
both summer and winter. Consistent with a strong covariance of precipitation and wind-driven
fluxes, Araligidad and Maloney (2008) showed a significant correlation between
intraseasonal TRMM precipitation anomalies and Tropical Atmosphere Ocean buoy latent
heat flux anomalies. For example, Figure~\ref{fig:scatter_lh_precip} 
is derived from Araligidad and Maloney (2008)
and shows a scatterplot of intraseasonal latent heat flux versus precipitation anomalies at
8S, 165E during November-April of 1999-2005, within the band of strongest
intraseasonal precipitation and OLR variance of Figures 2 and 3. If only the wind-driven
portion of the latent heat flux anomaly is retained in this analysis, the correlation is about
0.1 higher, indicating that intraseasonal anomalies in air-sea humidity difference (forced
primarily by SST variations) act to reduce the correlation of latent heat flux and
precipitation versus if the wind-driven component were acting alone.

A positive covariance between precipitation and latent heat flux is encouraging
regarding the ability for wind-evaporation feedbacks to support the MJO, in that it 
suggests that surface latent heat fluxes can influence convection.  However, for surface
flux feedbacks to drive the MJO, it is essential
is for wind-induced fluxes to engender a positive covariance of intraseasonal tropospheric
temperature and diabatic heating. Such a positive correlation would indicate
eddy available potential energy (EAPE) generation, with subsequent conversion of EAPE
to eddy kinetic energy supporting the large-scale MJO circulation against dissipation. It is difficult
to diagnose these energy conversions accurately from observations, but the evidence
from studies done to date suggests that the phase relationships are consistent with
a
role for surface enthalpy fluxes in the instability of the MJO.  
Hendon and Salby (1994)\nocite{hendon_salby_94} used satellite observations
of OLR and tropospheric temperature to show that heating and temperature are positively
correlated over the region of strong intraseasonal convective activity, being almost 
perfectly in phase in the Indian ocean where the MJO is growing in amplitude.  Similar
results were found in a more recent observational
study by Yanai et al. (2000)\nocite{yanai_et_al_00}.  Since
surface latent heat flux lags precipitation only by a small amount (compared to the 30-60
day period of the mode), and radiative heating is exactly in phase with precipitation (within
the accuracy of the observational estimates) the total surface enthalpy flux anomaly is
positioned to induce convective heating anomalies with the correct phase
to generate EAPE, particularly in the growing phase of the MJO life cycle.



\section{GCM results}
\label{sec:GCM}
\subsection{Previous work}

The hypothesis that interactive surface energy flux feedbacks are essential 
to the dynamics of intraseasonal variability
is testable in numerical models, under a perfect model assumption.  This can
be done by overriding the parameterizations which determine the net
surface enthalpy flux, or its individual components, and forcing the fluxes to
equal those from a climatology.  The climatological fluxes can be 
taken from a control run of the same model having interactive fluxes.  
Prescribing fluxes in this way renders the surface flux feedbacks inactive, as the surface fluxes in the
model are no longer a function of the instantaneous model variables.  If 
surface flux feedbacks are essential to the model's intraseasonal variability,
that variability should be eliminated, or at least significantly reduced in amplitude.
A number of variations on these experiments may be useful, such as one in 
which surface wind speed, rather than the turbulent surface fluxes themselves,
is prescribed.  Wind speed can either be set to a climatology or, in a simpler but 
less clean experiment shown below, to a spatially and temporally constant value.

To our knowledge, experiments of this type have been done only with a couple
of recent-generation general circulation models using realistic basic state SST. 
Maloney (2002)\nocite{mal02} performed an experiment in which the surface wind speed was set
to its climatological value in the computation of surface turbulent heat fluxes.  As 
intraseasonal variations in these fluxes are largely controlled by wind speed 
variations --- the WISHE feedback ---
this eliminated most (but not all) intraseasonal flux variations.  The surface latent heat
flux itself was set to its climatological seasonal cycle in one simulation in Maloney and
Sobel (2004).  In that study, eliminating surface flux feedbacks
significantly reduced the amplitude of the simulated MJO, 
indicating an important role for surface flux feedbacks.  The MJO was not totally eliminated,
but this could have been either due to fact that variations in
the other components of the net surface energy flux (the radiative fluxes, and the
turbulent flux variations due to air-sea humidity difference alone) were not suppressed, 
or due to dynamics unrelated to surface flux variations.  In Maloney (2002), on the 
other hand, the elimination of WISHE actually increased the amplitude of eastward-propagating
wind and precipitation variability.  The complete disagreement
between the results of these two studies is at first perplexing.
The models used in them were rather similar.  Both
used the relaxed Arakawa-Schubert convection parameterization in
successive versions of the NCAR Community Atmosphere model;  the two models
differed primarily in
the treatments of convective downdrafts and cloud microphysics. However, the resulting 
relationships between
intraseasonal convection and the large-scale anomalous circulation in these two models
was significantly different, with enhanced convection occurring in anomalous easterlies
and suppressed latent heat fluxes in the model of Maloney (2002), and in anomalous
westerlies and enhanced latent heat fluxes (as observed) in the study of Maloney and Sobel
(2004). Thus, removing wind-evaporation feedback might be expected to have different
effects in these two models. 

A few earlier GCM studies also tested the importance of surface turbulent
and cloud-radiative feedbacks to intraseasonal variability (e.g. Hayashi and Golder 1986) 
\nocite{hayashi_golder_86}, but given the considerable
advances in simulation capability in the last two decades, it may be 
most productive to focus on results from more recently developed models. 
In some relatively recent GCM studies using zonally-symmetric
SST distributions strong sensitivity to WISHE has been found (e.g. Hayashi and Golder
1997, Colon et al. 2002)\nocite{hayashi_golder_97}.  It might be argued that the 
differences between the basic state wind fields in these calculations and
the observed wind fields renders their relevance to real intraseasonal variability
somewhat indirect.  On the other hand, we do not understand that variability well
enough to be sure what the role of the basic state is.

\subsection{Results with the GFDL AM2}
\label{sec:am2}
In this section we present results from new 
simulations with the Geophysical Fluid Dynamics Laboratory's 
Atmospheric Model 2.1 (AM2.1), which is the atmospheric component of 
the coupled climate model CM2.1.  

\subsubsection{Model description}
With the exception of the modification 
to the convection scheme that we describe below, the model used here 
is identical to that presented by the GFDL Global Atmospheric Model 
Development Team \citep{andet04}.  It has a finite volume dynamical core, 
with $2^\circ \times 2.5^\circ$ horizontal resolution, and 24 vertical 
levels.  The model is run over realistic geography and climatologically
varying SSTs. The simulations are run for 11 years, with statistics taken 
over the last 10 years.

The convection scheme is a version of the Relaxed 
Arakawa-Schubert (RAS) scheme \citep{moosua92}.  In the RAS scheme, 
convection is represented by a spectrum of entraining plumes, 
with a separate plume corresponding to each model level that can be 
reached by convection.  The entrainment rates in these plumes are then 
determined by the requirement that the levels of neutral buoyancy of 
the plumes correspond to model levels.  The convection scheme 
in AM2 also uses the modification of \cite{toket88}, in which 
convection is not allowed to occur when the calculated 
entrainment rates are below a critical value $\lambda_0$ determined by 
the depth of the subcloud layer $z_M$, with $\lambda_0 = \alpha/z_M$.  
Thus with larger values of the Tokioka parameter $\alpha$, convection 
is prevented from reaching as deeply.  Inspired by the results of 
\cite{toket88} and \cite{linet08a} showing that larger values of the 
Tokioka parameter lead to stronger and more realistic MJO variability, 
we change $\alpha$ from its standard AM2 value of $\alpha = 0.025$ to 
the control value of $\alpha = 0.1$ in order to increase the MJO 
variance in the model.  We also show results using the 
standard AM2 value of the Tokioka parameter, in which the simulated
MJO is weak.

\begin{figure}
\noindent
\includegraphics[height=10cm,clip]{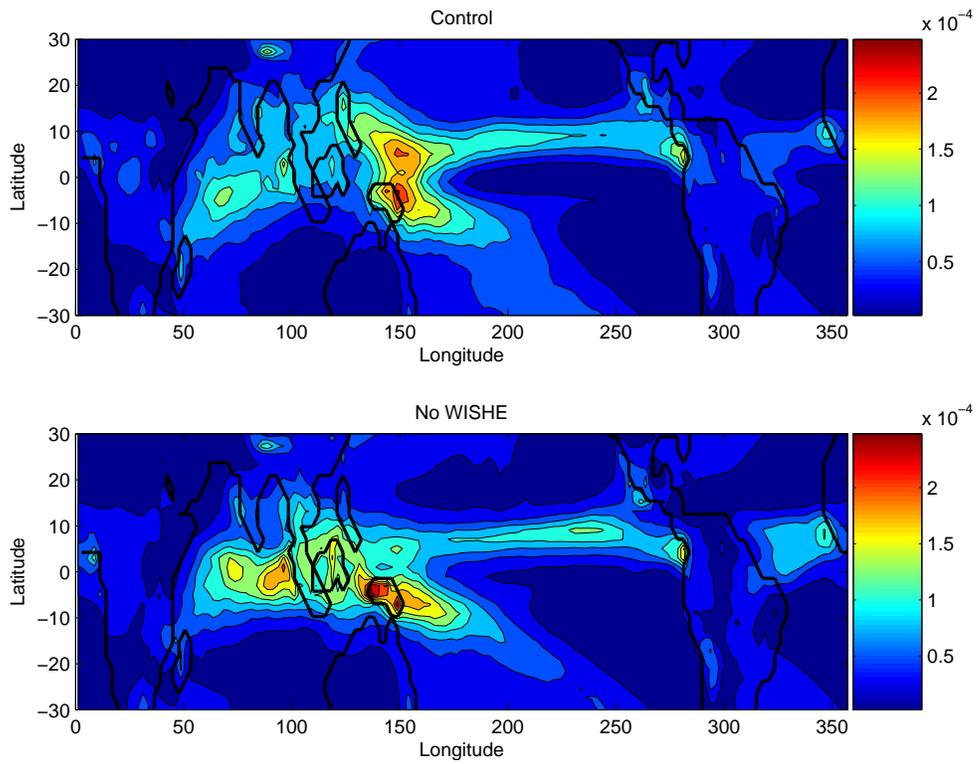}
\caption{Annual mean precipitation distributions 
for the control and no-WISHE simulations GFDL AM2 with Tokioka parameter $\alpha=0.1$
(see text for details).}
\label{fig:pannmean_tok}
\end{figure}

In order to identify the importance of WISHE to the model MJO, we construct
no-WISHE simulations by replacing 
the wind speed dependence in the surface flux formulation with a constant 
value representative of typical values over the tropics, $6\ m/s$ 
everywhere.  This modification is very simple to implement, 
as it only requires changing one line of code.  However, in addition to
preventing intraseasonal variations in turbulent surface fluxes (the intended
effect) this modification also alters the climatology of the model, since the model's
actual climatological surface wind speed is not constant in either space or season.  
We therefore must confirm 
that the climatological precipitation distribution in our no-WISHE simulation
does not become so different from that in the control model (or in observations)
as to render the experiment irrelevant.  The annual mean precipitation distributions for the 
control and no-WISHE cases are plotted in Figure \ref{fig:pannmean_tok}.  This 
figure shows that while there are some important changes in the 
precipitation distribution when WISHE is removed (e.g., less 
precipitation in the NW Pacific and more precipitation in the Indian 
Ocean), the distributions remain qualitatively similar enough to merit 
comparison of the MJO characteristics.  

\subsubsection{Results}

\begin{figure}
\noindent
\includegraphics[width=5cm]{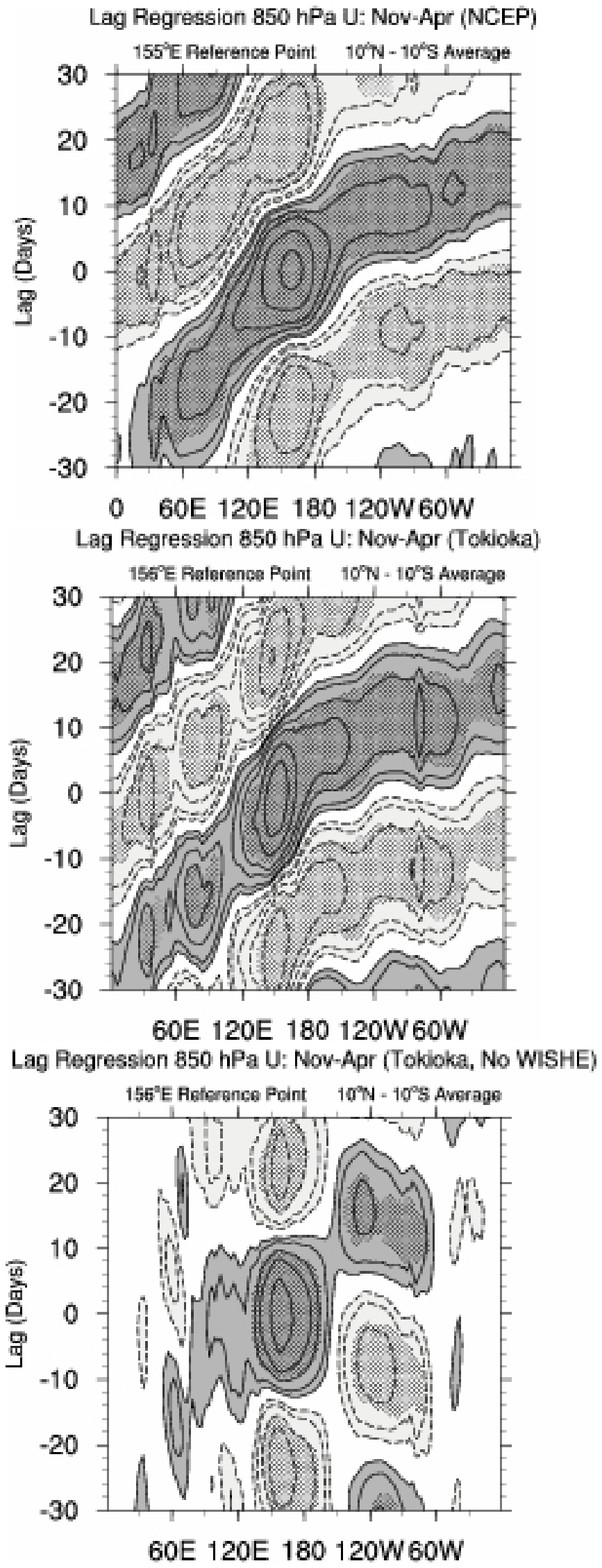}
\caption{Lag-correlation of $10^\circ S - 10^\circ N$ averaged, 30-90 day filtered
zonal wind at 850 hPa (U850) against the time series of the same field at $156^\circ E$,
in GFDL AM2 with Tokioka parameter $\alpha=0.1$.  The top panel shows results
from the NCEP/NCAR reanalysis, the middle shows results from the model, and
the bottom shows results from the model with no WISHE (see text for details).}
\label{fig:lagcorr_tok}
\end{figure}

As a first measure of the strength of the MJO with and without WISHE in 
these simulations, we show in Figure \ref{fig:lagcorr_tok} lag-regression 
plots for 30-90 day filtered equatorial ($10^\circ N-10^\circ S$ averaged) 
zonal wind at 850 $hPa$ (U850) for the months of 
November-April for observations, the control case (with $\alpha=0.1$), 
and the control case 
without WISHE.  Regression coefficients are scaled by the 1 sigma value of 
the reference $156^\circ E$ time series, and stippling indicates where the correlation 
coefficient is significantly different from zero at the 95\% confidence 
level.  The control case MJO propagation is quite similar to 
observations in many aspects, including implied phase speeds of 
approximately 5 $m/s$, large variance over the Indian and Pacific Ocean, 
and faster propagation over the central/eastern Pacific, though the regression 
coefficients are generally a bit weaker than those in observations, especially 
just to the west of the reference point.  When WISHE is suppressed, the 
amplitude of the MJO is reduced signficantly.  Only over the Pacific 
does any significant correlation exist away from the reference point.  
This clearly demonstrates that the MJO in this model is strongly 
influenced by WISHE.

As an alternative measure of the MJO intensity, we examine the 
intraseasonally averaged (30-90 days) wavenumber spectrum for U850, 
separated into eastward and westward propagating components.  The ratio 
of eastward to westward variance at wavenumber 1 is often used as a 
measure of the MJO strength (see, e.g., \cite{zhaet06}).  The ratio of 
eastward to westward variance at wavenumber 1 in the control case is 2.60, 
which is stronger than nearly all the atmosphere-only models in the 
\cite{zhaet06} study, although weaker than observations, which have a 
value of 3.5.  When WISHE is removed, this E/W ratio is reduced to 1.16, 
corroborating the result that WISHE is fundamentally important to the MJO 
in this model.  

\begin{figure}
\noindent
\includegraphics[width=5cm]{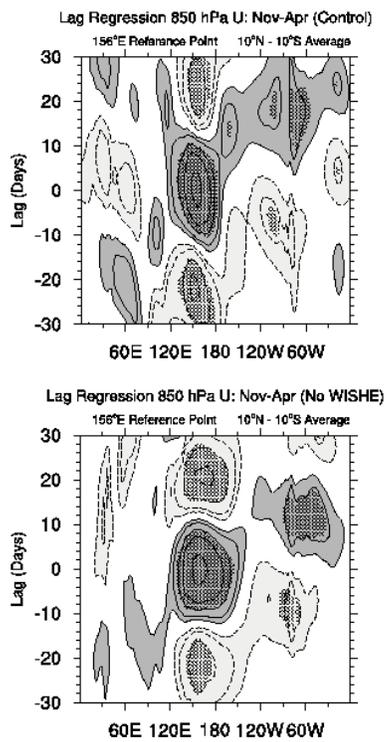}
\caption{Lag-correlation of U850 in GFDL AM2 , as in the lower two panels of
fig.~\protect\ref{fig:lagcorr_tok}, but
with Tokioka parameter $\alpha=0.025$.}
\label{fig:lagcorr_notok}
\end{figure}

As a test of robustness to changes in model physics, we examine the same 
MJO diagnostics for the standard version of AM2.1, in which the Tokioka 
parameter $\alpha = 0.025$.  
Examining lag correlations for this configuration in Figure 
\ref{fig:lagcorr_notok}a, one can clearly see weaker MJO propagation everywhere 
as compared to the Tokioka-modified control case in Figure 
\ref{fig:lagcorr_tok}b.  Correlations are significantly weaker, especially
in the Pacific basin.  When WISHE is removed in this model 
configuration (Figure \ref{fig:lagcorr_notok}b), the
model MJO is little affected.  There is a small indication of decreased MJO correlations in the 
Indian Ocean and immediately downstream of the reference point, but these 
changes are subtle.  The ratio of eastward to westward intraseasonal 
variance at wavenumber 1 for U850 is reduced from 1.31 to 1.08 when WISHE 
is removed, indicating a small decrease in MJO amplitude with 
WISHE in this diagnostic.  Generally speaking, removing WISHE has a small 
effect on the MJO in this model configuration;  but then, the MJO is weak to
begin with.  

\clearpage 

\clearpage

The results from the model with
$\alpha=0.1$ bring to two the number of recent-generation
models in which WISHE has been found to be important to the MJO in 
simulations with realistic basic states, the other being that
used by Maloney and Sobel (2004).  Both models use versions of the
RAS convective parameterization, so they are not entirely unrelated,
but most other aspects of the two models are different.  In the models of
Maloney (2002) and the AM2.1 with $\alpha=0.025$, WISHE is not
important to the simulated MJO.  On the other hand, these two models
have MJO simulations which resemble observations less closely than 
do those of Maloney and Sobel (2004) and the AM2.1 simulations with
$\alpha=0.1$.  At least for this small sample of models, a better simulation
of the MJO seems to be associated with an increased role for WISHE.

The simulations discussed above address only the role of the surface
latent heat flux, and in the case of Maloney (2002) and the AM2.1 calculations
described here, only the wind-induced component of that flux.  A few ,
studies with full-physics GCMs over realistic 
of continents and sea surface temperature have assessed the role of radiative
flux perturbations in simulated MJO dynamics.  
Lee et al. (2001)\nocite{leeet01} found in an aqua-planet GCM that 
overactive cloud-radiative feedbacks degraded the MJO simulation by inducing spurious
small-scale disturbances;  reasonable
changes to the model's physical parameterizations mitigated this degradation.
Other studies have been done in more idealized frameworks.
Raymond (2001)\nocite{ray01} 
found that cloud-radiative feedbacks were essential to the MJO simulated in his 
intermediate-complexity model, with surface turbulent fluxes also playing
a significant role.  
Grabowski (2003)\nocite{grabowski_03} found in aqua-planet simulations with
what is now called the "multiscale modeling framework" (MMF) that cloud-radiative
feedbacks were not important to his simulated MJO disturbances, while surface
latent flux feedbacks were essential to the disturbances' development.
Lin et al. (2008)~\nocite{lin_et_al_08} found no effect of cloud-radiative feedbacks on the
MJO in a model in which the MJO was weak to begin with.
 
\section{Discussion}
\label{sec:discussion}
\subsection{The crux of the matter, in theoretical context}

The claim that surface enthalpy fluxes are essential to the dynamics of tropical
intraseasonal variability is not new, going back at least 20 years to the studies
of Emanuel (1987) and Neelin et al. (1987).  We believe that, given the lack of
broad agreement on the mechanisms of the MJO (despite decades of intense
effort) and the evidence from both observations and GCMs discussed above
to support the hypothesis that surface fluxes are important, the time has come 
to reassess this hypothesis in a more focused way.  

We have not provided a comprehensive discussion of all theories for
tropical intraseasonal variability.   However, it should be uncontroversial to state
that in many of these theories, interactive surface fluxes either are not 
essential or are absent altogether.  
We propose that it would be useful to divide the current set of theories into two 
subsets, one in which feedbacks involving 
surface fluxes (including radiative fluxes) are essential and one in which they are not,
and then attempt to eliminate one subset via focused numerical modeling studies, perhaps
combined with further analysis of observations.

From a purely conceptual point of view, whether surface fluxes are essential to
intraseasonal variability is a fundamental question.  In extratropical dynamics, it has been found
useful to divide the set of possible dynamical processes into those which are
dry adiabatic and those which are not.  If a phenomenon can be understood using
adiabatic models, it is advantageous to do so.  Similarly, it is natural when discussing 
tropical dynamics to divide the large set of possible processes
into those which involve only deep convection and large-scale dynamics --- that is,
those which can be represented by moist adiabatic dynamics --- and those in which 
diabatic processes external to both
deep convection and large-scale dynamics, namely turbulent surface fluxes
and radiative cooling, are involved.  It has been a goal of theoretical tropical
meteorology for several decades to determine whether the interaction of 
convection and large-scale dynamics alone can generate large-scale
variability [as in early CISK models, as well as in more recent models with
more complex physics (e.g., Mapes 2000; Majda and Shefter 2001;  Kuang 2008)]
\nocite{map00,majshe01a,khomaj06a,khomaj06d,kua08b}
or whether interaction with diabatic processes external to
convection and large-scale dynamics is necessary (as in first baroclinic mode
QE models).  Determining whether interaction with turbulent surface fluxes and
radiation is essential to observed intraseasonal variability in particular
would be a major step forward in our understanding of the tropical atmosphere.

\subsection{Proposal for further model intercomparison}

Recent model intercomparisons~\citep{zhaet06,linet06}, have been performed which summarize
the state of the art in simulating the MJO in general circulation models.  Without restating
the results of these studies in detail, the MJO simulations in the latest generation of models are 
on average superior
to those in previous generations~\citep{Slingo-Sperber-Boyle-et-al-1996:intraseasonal,Sperber-Slingo-Inness-et-al-1997:maintenance} in simulating eastward-propagating zonal wind 
variability in the tropics with a dominance of eastward vs. westward power,
though even the best models still have deficiencies in their MJO simulations.  
While keeping model deficiencies in mind, we 
propose that interested modeling groups perform experiments
like those described above, in which the total surface enthalpy flux, and ideally also its 
individual components, are set to climatology, eliminating feedbacks involving those fluxes.
These experiments are likely to yield unambiguous 
information about the dynamics of a model's intraseasonal variability.  The negative of 
the quantitative change in the strength of the intraseasonal variability in these experiments 
provides a direct estimate of the role of the eliminated feedbacks in the dynamics of the 
simulated variability in the control simulation.   
Besides GCMs with parameterized physics, these experiments
can also be done in models with resolved convection such as the multiscale modeling 
framework \citep{randall_et_al_03,khairoutdinov_et_al_05,khairoutdinov_et_al_08} or
global cloud resolving models (Miura et al. 2007)\nocite{miura_et_al_07}.  
Both of these technologies are showing
great promise in simulating the MJO, and sensitivity experiments to determine the roles
of surface turbulent and radiative flux feedbacks in their results would be particularly
valuable.

Because simulations of intraseasonal variability are imperfect in all models, such experiments
will not yield unambiguous information about the dynamics of intraseasonal variability in
the real atmosphere.  It is entirely possible that any given model, or even an entire generation
of models (given the broad similarities of approach found in common physical parameterizations
in climate models), is getting something close to the right answer for the wrong reasons, so that the
results of these experiments would be misleading.  It is perhaps also equally probable that 
different models will yield different results from these experiments. 

Neither will such experiments provide any direct information about how to improve the
simulation of intraseasonal variability in any given model.  The importance of
surface flux feedbacks  to the dynamics of intraseasonal variability may not be related 
in any simple way to any particular property of the physical parameterizations
of a model, nor to any other aspect of its construction (e.g., resolution or the dynamical
core).  These feedbacks are arguably
a high-level, or "emergent" property of a given model.  Even in the relatively simple models
used in theoretical studies, it is often not apparent what determines the importance of
surface flux feedbacks.  We might expect it to be even less obvious in comprehensive GCMs.

If we were fortunate, the importance of surface flux feedbacks in a model would be
related to that model's ability to simulate intraseasonal variability, as is the case in
the very small sample of models discussed in section 4.
We can imagine 
a scatter plot in which one axis is the fidelity of modeled intraseasonal variability to that observed
(how to quantify this is a separate problem which we do not address here);  the other axis is the importance of surface flux feedbacks
to intraseasonal variability
in the same model, as quantified by minus the change in MJO amplitude in an experiment
where the surface flux (either total, or a given component) is set to climatology;  and each
point represents one model.  A significant positive slope to the best-fit regression
line would suggest that surface flux feedbacks are important to dynamics of intraseasonal
variability in the real atmosphere, while a significant negative slope would suggest
the opposite.  Lack of any significant slope, of course, would be an ambiguous result.

In any case, knowledge of the role of surface fluxes in simulated intraseasonal variability
would be useful to model developers.  It seems
likely that any increase in physical understanding of the dynamics of the modeled intraseasonal
variability, such as quantification of the role of surface flux feedbacks, would help
 to guide in the formulation of hypotheses about how to improve a model.
For example, an active role for surface fluxes in regulating intraseasonal variability may compel modelers to further develop
parameterizations coupling mesoscale perturbations of moist entropy and gustiness to the
boundary layer (e.g. Jabouille et al. 1996;  Redelsperger et al. 2000)\nocite{jabouille_et_al_96}, 
where they may significantly affect
surface fluxes during MJO events.

\subsection{Theoretical challenges}

A determination that interactive surface fluxes are essential to 
the dynamics of intraseasonal variability would not constitute a complete
theory for that variability.  Even if we were able to resolve
the importance of surface fluxes, questions that would remain
unanswered include (among others):  What is the relative importance
of turbulent vs. radiative fluxes?  How should the physics of deep convection and other
unresolved processes be parameterized
in order to yield the correct feedback between the fluxes and the large-scale dynamics of
the mode?  Are the large-scale dynamics essentially linear or nonlinear?  Is the structure
of the basic state critical?  What sets the phase speed of the disturbances?  What is
the role of ocean coupling?  Theorists
currently struggle with all of these questions.  Proving or disproving
the hypothesis that surface fluxes are essential to tropical intraseasonal variability would 
tell us that the ultimate energy source for
the disturbances is the ocean mixed layer, and in doing so would eliminate a large
subset of theories, but much theoretical work would
be left to do.

Assuming that the wind-induced component of the surface turbulent fluxes is important, 
the dynamics by which these fluxes interact with the dynamics of eastward-propagating
MJO disturbances must be different than that envisioned by Neelin et al. (1987) and Emanuel (1987).  
The real MJO is not a Kelvin wave (though it retains aspects of Kelvin wave dynamics), and the basic state surface winds in regions of strong tropical intraseasonal variability are westerly (e.g. Inness and Slingo 2003;  Maloney and Esbensen 2007)\nocite{inness_slingo_03,maloney_esbensen_07}.
One possibility is that nonlinear WISHE, rather than its linear counterpart, is acting.  
Studies which present numerical simulations with
idealized or intermediate-complexity models~\citep{ray01,sugiyama_08a,sugiyama_08b} 
as well as comprehensive models (Maloney and Sobel 2004) 
provide suggestions of how this might work, but
we do not have a simple analytical prototype model for nonlinear WISHE.
Another possibility is that the dynamics are still fundamentally linear, but that fundamental
changes to other aspects of the original E87 and N87 models 
(e.g., the identification of the MJO as a Kelvin wave)allow the requirement of
mean easterlies to be relaxed. In the idealized model of Wang and Xie (1998)\nocite{wang_xie_98}, 
for example, the combination of coupling to a mixed layer ocean and parameterized radiative feedbacks
is linearly destabilizing in the presence of mean westerlies. 

In the case of the northern summer northward-propagating mode, there may be no
fundamental theoretical problem. In at least one idealized model, a mode resembling that
observed is linearly destabilized by WISHE (Bellon and Sobel 2008a, b).

\section{Summary}

We have argued that feedbacks involving the total surface enthalpy flux are important
to the dynamics of tropical intraseasonal variability, possibly providing the primary 
energy source for intraseasonal disturbances.  Observational evidence presented in support of
this argument consists of maps of intraseasonal variance in precipitation and OLR as well
as local correlations between precipitation and surface fluxes on intraseasonal time scales.
Modeling evidence consists of results from several GCMs as well as idealized models in
which surface flux feedbacks are demonstrably important if not essential to simulated 
intraseasonal variability.

Our understanding of the mechanisms responsible for
 intraseasonal variability is still poor after decades of study.  We have argued that
 the time has come for a more systematic evaluation of the role of surface enthalpy
 fluxes, given all the tools at hand, with the aim of eliminating from consideration
 either those hypotheses in which surface fluxes are important or those in which they
 are not.  Given the evidence presented here, the increasing fidelity
with which comprehensive numerical models simulate intraseasonal variability, and the relative 
straightforwardness of assessing
 the importance of surface flux feedbacks in those models, we have argued that it would
be particularly useful if a larger number of interested modeling groups were to perform the 
necessary assessments.  Such efforts, combined with targeted 
observational and theoretical work, might enable the field to move forward in a more
coordinated and productive way towards a better understanding of tropical intraseasonal
variability.

 \section*{Acknowledgments}
This paper was written while the first author spent a sabbatical year as a visitor 
at the Centre for Australian Climate and Weather Research (CAWCR), Bureau of Meteorology
(BoM), Melbourne, Australia.  Some of the writing was done during a relatively brief but very stimulating visit (during the peak of an active MJO event) to the BoM's Northern Territory Regional Office in Darwin.
He (AHS) is grateful to the BoM for its hospitality
during this year, and to its scientists Harry Hendon, Matt Wheeler, and Hongyan Zhu (Melbourne) and
Sam Cleland and Lori Chappell (Darwin) for stimulating discussions on the
topic of this paper.  This work was
supported by the Climate and Large-Scale Dynamics Program of the National
Science Foundation under grants ATM-0632341 (EDM) and ATM-054273 (AHS), 
by the Precipitation Measurement Mission program of the National Aeronautics
and Space Administration under grant NNX07AD21G (AHS) and by award
NA05OAR4310006 from the National Oceanic and Atmospheric Administration,
U.S. Department of Commerce (EDM). The statements, findings, conclusions, and
recommendations do not necessarily reflect the views of NSF, NASA, NOAA, or of the
Department of Commerce.

\bibliographystyle{./jaslike}
\bibliography{./lit,all,bibmjo}

\end{document}